\documentclass[10pt,twocolumn,letterpaper]{article}

\usepackage{times}
\usepackage{multirow} 
\usepackage{epsfig}
\usepackage{graphicx}
\usepackage{amsmath}
\usepackage{amssymb}
\usepackage{subcaption}
\usepackage{overpic}
\usepackage{wrapfig}
\usepackage{booktabs}
\usepackage{verbatim}
\usepackage[accsupp]{axessibility}
\usepackage[table]{xcolor} 
\usepackage{colortbl}

\usepackage{cvpr}              
\usepackage{bibunits}
\defaultbibliographystyle{ieeenat_fullname}
\defaultbibliography{main}

\definecolor{cvprblue}{rgb}{0.21,0.49,0.74}
\usepackage[breaklinks,colorlinks,allcolors=cvprblue]{hyperref}

\def\paperID{31716} 
\def\confName{CVPR}
\def\confYear{2026}

\title{ChronoGS: Disentangling Invariants and Changes in Multi-Period Scenes}

\author{
Zhongtao Wang\textsuperscript{1} \quad
Jiaqi Dai\textsuperscript{2} \quad
Qingtian Zhu\textsuperscript{3} \quad
Yilong Li\textsuperscript{1} \quad
Mai Su\textsuperscript{1} \\
Fei Zhu\textsuperscript{1}\footnotemark[1] \quad
Meng Gai\textsuperscript{1} \quad
Shaorong Wang\textsuperscript{2}  \quad
Chengwei Pan\textsuperscript{4} \quad
Yisong Chen\textsuperscript{1}\footnotemark[1] \quad
Guoping Wang\textsuperscript{1} \quad \\  \\
\textsuperscript{1} Peking University \quad 
\textsuperscript{2} Beijing Forestry University \quad 
\textsuperscript{3} The University of Tokyo \quad 
\textsuperscript{4} Beihang University
}

\begin{document}

\twocolumn[{%
\renewcommand\twocolumn[1][]{#1}%
\maketitle
\vspace{-1.5em}
   \centering
   \includegraphics[width=.96\textwidth]{figs/new_teaser_down.pdf}
\captionof{figure}{As time goes by, scenes naturally evolve — sunlight shifts, new structures emerge, seasons change, and so on. Cross-period captures therefore exhibit variations in both geometry and appearance. The \textcolor{red}{red} and \textcolor{blue}{blue} overlays in the illustration simply exemplify the differences between the two periods. These coupled changes pose significant challenges for existing methods. Our approach, \textbf{ChronoGS}, robustly reconstructs and disentangles both geometry and appearance variations, delivering consistent and faithful results across time.
\vspace{2em}}
\label{fig:teaser}
}]

\footnotetext[1]{Corresponding authors. Yisong Chen is the lead corresponding author.}

\begin{bibunit}

\begin{abstract}
Multi-period image collections are common in real-world applications.
Cities are re-scanned for mapping, construction sites are revisited for progress tracking, and natural regions are monitored for environmental change.
Such data form multi-period scenes, where geometry and appearance evolve. Reconstructing such scenes is an important yet underexplored problem.
Existing pipelines rely on incompatible assumptions: static and in-the-wild methods enforce a single geometry, while dynamic ones assume smooth motion, both failing under long-term, discontinuous changes. To solve this problem, we introduce \textbf{ChronoGS}, a temporally modulated Gaussian representation that reconstructs all periods within a unified anchor scaffold. It's also designed to disentangle stable and evolving components, achieving temporally consistent reconstruction of multi-period scenes.
 To catalyze relevant research, we release \textbf{ChronoScene} dataset, a benchmark of real and synthetic multi-period scenes, capturing geometric and appearance variation. Experiments demonstrate that ChronoGS consistently outperforms baselines in reconstruction quality and temporal consistency.
 Our code and the ChronoScene dataset are publicly available at \hyperlink{https://github.com/ZhongtaoWang/ChronoGS}{https://github.com/ZhongtaoWang/ChronoGS}.
\end{abstract}

\section{Introduction}
\label{sec:intro}

Urban environments are periodically re-scanned for city mapping and management. Active construction sites are repeatedly surveyed to monitor progress and structural modifications. Regions affected by natural disasters are revisited over time to assess damage and document recovery. As a result, multi-period imagery of the same scene naturally accumulates—collected under different seasons, lighting, and even structural layouts. These multi-period captures contain valuable signals for long-term scene understanding, yet also bring unique challenges: geometry and appearance may evolve discontinuously due to construction, vegetation growth, or environmental variation.

Despite its clear importance, this multi-period scene reconstruction setting has remained largely unexplored. Existing pipelines are typically designed for two distinct settings. Static methods, including those designed for in-the-wild imagery, assume that all views share a time-invariant geometry. Dynamic methods, on the other hand, rely on smooth and continuous motion over time. Neither assumption holds when captures are seasons or even years apart. Consequently, current approaches fail to provide a consistent representation while flexibly modeling per-period variations. Our work aims to close this gap by introducing a framework that reconstructs all periods within a unified representation, disentangling what remains stable from what evolves.

We define a multi-period\footnote{We use the term \textbf{period} rather than \textbf{timestamp}. Because in our setting, each period corresponds to a time window during which images are captured at slightly different times. We manually define periods using metadata or prior knowledge, assuming that changes within a period are small compared to those across periods.} scene as a set of captures of the same spatial region acquired at discrete time intervals, where the scene may have undergone geometric or appearance changes between captures. Our key observation is that multi-period scene reconstruction is neither static nor smoothly dynamic — it is discrete in periods but shared in major structure. The majority of spatial content remains invariant across periods, while changes are independent, enabling the scene to be decoupleable. This property allows the scene to be factorized into a shared canonical geometry and period-specific variations, which can be independently modeled within a unified framework.

Therefore, we propose \textbf{ChronoGS}, a temporally modulated Gaussian representation that models multi-period scenes consistently. 
ChronoGS reconstructs all multi-period scenes within a unified anchor scaffold of union geometry across all periods, providing a consistent geometric backbone for cross-period reconstruction. 
Within this scaffold, each anchor maintains an invariant base feature and a pool of local period-varying features that capture temporal variations. In addition, a global period-varying feature is adopted to model scene-level changes. These features allow anchors to vary over time while remaining a unified structure, ensuring stable and consistent reconstructions across periods.
A geometry activation mechanism further enables the scaffold to express geometry specific to each period while maintaining global spatial consistency. 
Through this unified design, ChronoGS captures complex temporal phenomena, including both geometry and appearance change, 
achieving consistent reconstruction on multi-period scenes.

To support our research, we introduce \textbf{ChronoScene}, 
a novel benchmark designed to capture both geometry and appearance changes of multi-period scenes. 
Unlike existing benchmarks that assume static geometry or short-term continuous motion, this setting exposes the real challenges of large-scale scene evolution—building construction, vegetation growth, seasonal, and lighting shifts, etc.
ChronoScene aligns multi-period captures in a common world coordinate frame, and provides a principled testbed for temporally modulated 3D modeling approaches.

Our experiments demonstrate that ChronoGS outperforms existing methods across diverse scenes, delivering better renders and more accurate geometry for each period while synthesizing plausible intermediate states. Ablation studies prove the rationality of our design. Also, we hope that ChronoGS and our dataset establish a foundation for future research on multi-period scene understanding.

\noindent\textbf{Our contributions are threefold:}
\begin{itemize}
\item \textbf{Novelty.} To the best of our knowledge, this is the \textbf{first} work to reconstruct multi-period scenes within a unified, differentiable Gaussian-splatting framework, jointly handling non-continuous geometry and appearance changes.

\item \textbf{Dataset.} We introduce the \textbf{ChronoScene} dataset, a novel benchmark that aligns multi-period observations within a common coordinate and captures both geometric and appearance evolution across periods.

\item \textbf{Method.} We propose \textbf{ChronoGS}, a temporally modulated Gaussian representation that effectively handles geometric and appearance variations that exist in multi-period scenes.
\end{itemize}

\section{Related Work}
\label{sec:related_work}

\subsection{Neural Radiance Fields}

Neural rendering has advanced along two complementary tracks. Implicit approaches such as NeRFs~\cite{mildenhall2021nerf,barron2022mip,reiser2023merf,pumarola2021d,tancik2023nerfstudio} represent scenes as continuous volumetric fields, mapping 3D location and viewing direction to density and radiance via learned networks coupled with volumetric rendering. While achieving photorealistic novel views, dense ray sampling and repeated network evaluations are computationally expensive, prompting extensive work on acceleration~\cite{muller2022instant,barron2022mip,fridovich2022plenoxels}. In parallel, explicit point-based methods target real-time performance: 3D Gaussian Splatting~\cite{kerbl20233d} models scenes with rasterized Gaussian primitives for high-quality rendering, extended by hierarchical organization in Scaffold-GS~\cite{lu2024scaffold}, scalable variants for large scenes~\cite{ren2024octree,kerbl2024hierarchical,su2025hug}, and extract view-consistent surface representations~\cite{huang20242d,chen2024pgsr}.

Despite these advances, both paradigms assume static scenes, forcing a single time-invariant representation of all observations. Jointly training all periods causes temporal averaging and artifacts, whereas per-period training avoids interference at the cost of heavier computation and storage without improving results. Our ChronoGS enables joint optimization across multiple periods, maintaining consistent long-term reconstruction in a unified model.

\begin{figure*}[ht]
    \centering
    \includegraphics[width=\textwidth]{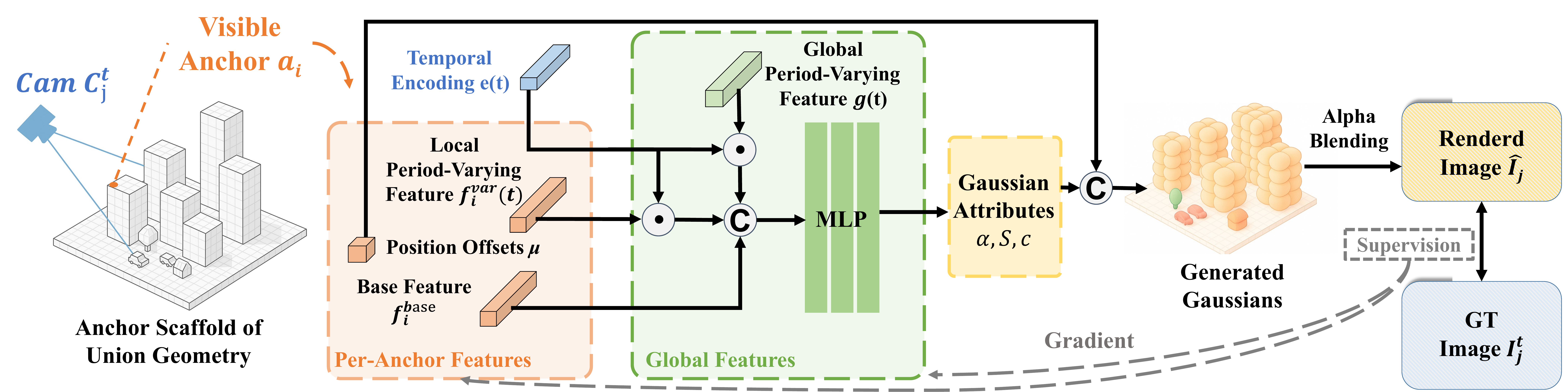}
    \vspace{-1.5em}
    \caption{
For a given camera $C_j^t$, we select visible anchors from our learned anchor scaffold of union geometry across periods.  We use per-anchor's features and global features to render the multi-period scene. 
Temporal behavior is modeled via the anchor's local period-varying feature $f_i^{\text{var}}(t)$ and global period-varying feature $g(t)$ modulated by temporal encoding $e(t)$. After concatenating them with anchor's period-invariant base feature $f_i^{\text{base}}$, a lightweight MLP predicts Gaussian attributes $\{\alpha, S ,c\}$ . These attributes, along with the learned Gaussian position offset $\mu$, produce a small cluster of Gaussians per anchor. 
We render the Gaussians generated by all visible anchors via differentiable splatting with alpha blending to obtain the image $\hat{I}_j^t$, 
and optimize with photometric losses between the ground truth image. 
This design enables consistent reconstruction of both geometry and appearance evolution on multi-period scenes.
}
    \label{fig:pipeline}
\end{figure*}

\subsection{Dynamic Scene Representation}

Building upon the success of NeRF in static scenes, several works have focused on extending it to dynamic environments. D-NeRF~\cite{pumarola2021d} models motion by constructing a canonical space and applying deformation MLPs, although they are constrained by long-sequence videos. K-Planes~\cite{fridovich2023k} decompose features across multiple planes, enabling lightweight models to capture motion effectively. In parallel, 3DGS\cite{kerbl20233d} has been extended to dynamic scenes. Dynamic 3DGS~\cite{luiten2024dynamic} optimizes Gaussians shared across time with regularizers, which rely on accurate point-cloud segmentation and additional temporal consistency terms. 4DGS~\cite{wu20244d} models dynamic scenes using canonical-space Gaussians combined with deformation MLPs. \cite{yang2023real,duan20244d} lift 3D Gaussians into 4D by adding a temporal dimension, and extract the corresponding 3D Gaussians at a given timestamp for rendering. 

However, many of these methods assume temporally dense and smooth evolution, depend on precise masks or segmentations, and struggle with abrupt appearances and disappearances. These limitations are especially acute in multi-period data with long gaps and discontinuities. In contrast, our ChronoGS framework decouples the representation over periods, accommodating sudden shifts while preserving cross-period consistency.

\subsection{Long-Term Scene Evolution Modeling}

Several prior works study temporal evolution in visual data, but their goals differ from ours. Early approaches~\cite{schindler2007inferring,schindler2010probabilistic,matzen2014scene} infer temporal order from opportunistically captured Internet photos, while~\cite{schindler20124d} emphasizes organization and visualization of historical photos with temporal metadata. Incremental reconstruction and time-lapse methods~\cite{ulusoy2014image,martin20153d,martin2015time} build per-time updates or visualization outputs, rather than learning a unified renderable representation across periods.

Related \emph{in-the-wild} novel view synthesis methods improve robustness to uncontrolled captures (unknown poses, illumination shifts, post-processing, and transient occluders). NeRF-W~\cite{martin2021nerf} models per-image appearance and transients, and NeRF-MS~\cite{li2023nerf} regularizes appearance codes for cross-sequence consistency. NeuSC~\cite{lin2023neural} addresses frequent appearance changes but assumes fixed geometry. In explicit representations, in-the-wild 3DGS methods~\cite{kulhanek2024wildgaussians,xu2024wild,lin2024vastgaussian,zhang2024GS-W,wang2024we} mainly target robust reconstruction under appearance variation and incremental updates. Continual adaptation methods such as \cite{cai2023clnerf,ackermann2025cl,zeng2025gaussianupdate} mitigate forgetting under incremental scene changes, but primarily capture relatively local and gradual geometry or appearance variation.

Overall, existing methods mostly focus on timeline inference, visualization, appearance variability, or incremental updating, and often collapse observations into a single state or assume fixed geometry. ChronoGS instead targets discontinuous multi-period captures with significant changes in both geometry and appearance, and learns a unified time-modulated representation that preserves cross-period consistency while enabling faithful period-specific rendering.

\section{Method}

Given a multi-period scene's captures $\{(\mathbf{I}^{(t)}_j, \mathbf{C}^{(t)}_j)\}_{j=1}^{N}$ containing $N$ images $\mathbf{I}^{(t)}_j$ and camera parameters $\mathbf{C}^{(t)}_j$ collected at $t=1 \ldots T$ discrete periods, our goal is to reconstruct all periods within a unified 3D representation. 
Different periods are often collected months or years apart, leading to substantial changes in both geometry and appearance.

\begin{figure}[!t]
    \centering
    \includegraphics[width=\linewidth]{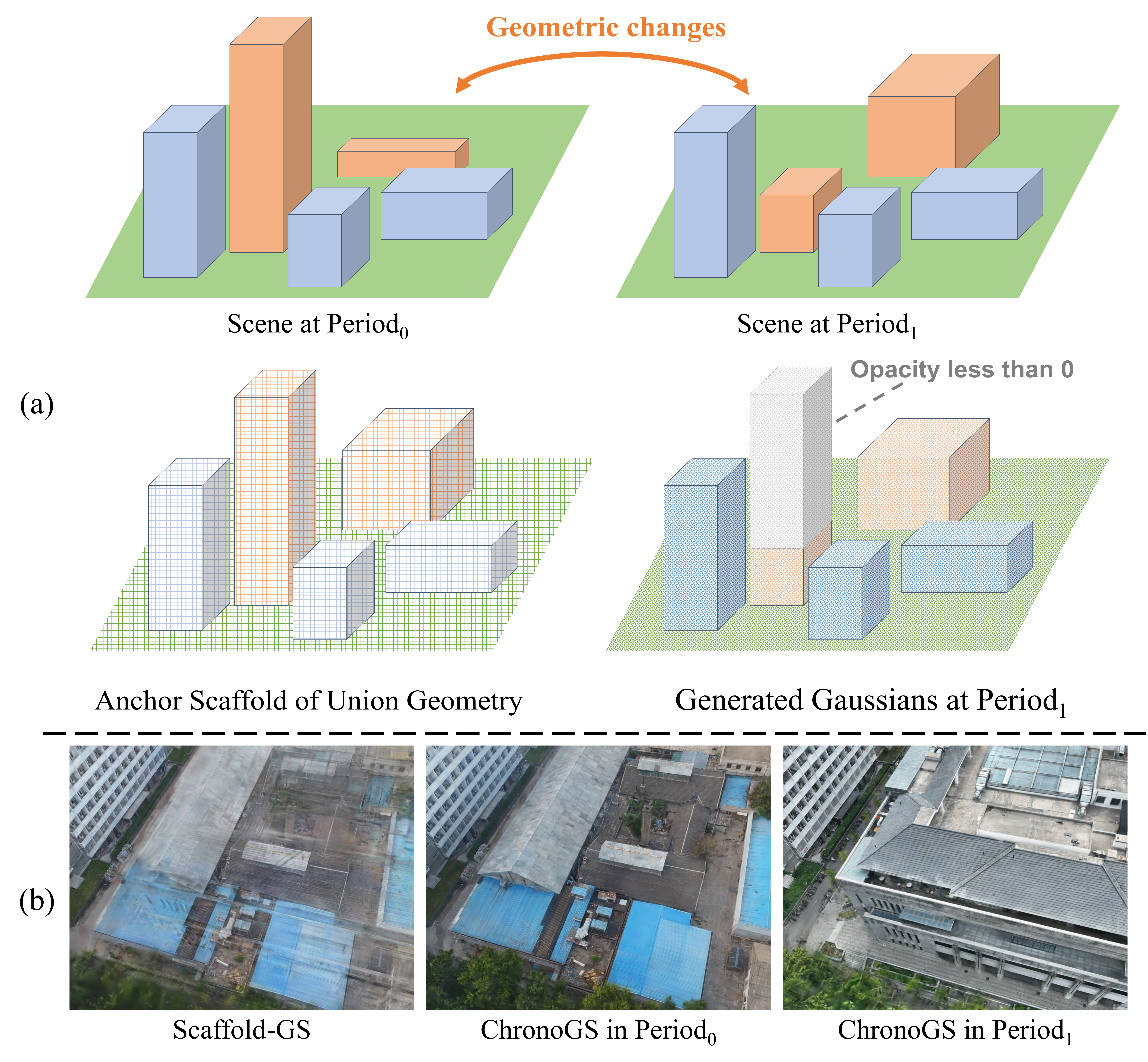}
    \vspace{-1em}
    \caption{\textbf{Illustration of the temporal geometry activation mechanism.}
\textbf{(a).} Anchor scaffold encodes the union of geometry across all periods, and generated gaussians with negative opacity are deactivated at certain periods, enabling the model to adaptively represent geometry variations over periods. 
\textbf{(b).} Comparison between Scaffold-GS and our ChronoGS trained on all periods' images. Scaffold-GS entangles structures from different periods, while ChronoGS successfully distinguishes and reconstructs period-specific geometry at $\text{Period}_0$ and $\text{Period}_1$.
}
\vspace{-1.5em}
    \label{fig:geo_filter}
\end{figure}

We propose a temporally modulated Gaussian representation that can automatically distinguish period-varying and invariant components.
Following the anchor-based paradigm of Scaffold-GS~\cite{lu2024scaffold}, we represent the scene with a set of \emph{anchors} $\mathcal{A}=\{a_i\}_{i=1}^{N_a}$, each located at a fixed 3D position $\mathbf{x}_i$ and carrying learnable feature vectors.
Given a camera, we select the visible anchors via frustum culling, and a lightweight MLP decodes each anchor's features into a small cluster of $K$ 3D Gaussian primitives, which are used for differentiable splatting.
The scaffold is initialized by voxelizing the merged sparse point clouds from all periods into a uniform spatial grid, so that both stable and varying geometry are covered from the start.
Rather than training separate models per period, we optimize a single scaffold across periods, while temporal modulation and a geometry activation mechanism identify regions that change in appearance or geometry, enabling faithful per-period reconstruction and plausible intermediate-state synthesis. An overview is illustrated in \cref{fig:pipeline}.

\subsection{Temporally Modulated Feature Representation}
\label{subsec:temporal_features}

\paragraph{Anchor feature decomposition.}
To jointly represent shared and period-specific properties, each anchor $a_i$ maintains two features:
a period-invariant base feature $\mathbf{f}^{\text{base}}_i \in \mathbb{R}^{d_b}$ that describes shared geometry and appearance,
and a local period-varying feature pool 
$\mathbf{f}^{var}_i=[\mathbf{f}^{(1)}_i,\dots,\mathbf{f}^{(T)}_i]\in\mathbb{R}^{T\times d_v}$ that stores period-specific information.

\paragraph{Temporal Encoding.}
We represent temporal states using a unified $T$-dimensional encoding defined over period indices.
For integer-valued period indices, we adopt a one-hot encoding where each observed period corresponds to a unique basis vector.
For the inter-period position, we interpolate between the embeddings of adjacent periods to obtain a smooth representation:
\begin{equation}
\label{eq:time-encoding}
\mathbf{e}(t) = (1-w)\mathbf{e}_{\lfloor t \rfloor} + w\mathbf{e}_{\lceil t \rceil}, \quad
w = t - \lfloor t \rfloor,
\end{equation}
where $\mathbf{e}_{\lfloor t \rfloor}$ and $\mathbf{e}_{\lceil t \rceil}$ are the one-hot embeddings of neighboring periods.
This formulation preserves exact encodings for observed periods while enabling smooth interpolation between adjacent periods.

\paragraph{Global–Local Temporal Modulation.}
The temporal encoding $\mathbf{e}(t)$ defined in \cref{eq:time-encoding} modulates both the local period-varying feature $\mathbf{f}^{\text{var}}_i(t)$, capturing per-anchor temporal variations,
and a global period-varying feature $\mathbf{g}(t)\in\mathbb{R}^{T \times d_g}$ shared across all anchors, capturing scene-level factors such as illumination and seasonal appearance.
Together with the period-invariant base feature $\mathbf{f}^{\text{base}}_i$, these three components allow the model to capture fine-grained dynamics while maintaining temporal consistency.

\paragraph{Explicit Gaussian Offsets for Geometric Stability.}
The $i$-th anchor stores $K$ learnable but period-invariant Gaussian center offsets $\{\Delta\boldsymbol{\mu}_{ik}\}_{k=1}^{K}$, 
forming a fixed local arrangement around the anchor. 
Explicitly storing these offsets preserves local geometric stability across periods, 
so temporal geometry changes are mainly reflected through the Gaussians' visibility.
This fixed spatial layout ensures consistent geometry across periods, enabling reliable opacity-based filtering described in \cref{subsec:anchor_scaffold}.


\subsection{Modeling Cross-Period Unified Geometry}
\label{subsec:anchor_scaffold}

\paragraph{Temporal geometry activation.}
As a unified framework, we design the anchor scaffold to align with the union geometry across all periods of the scene, and let the model learn which anchors should be activated at different periods. We model geometry as locally stable around each anchor, while temporal changes are captured by modulating their visibility.
As illustrated in \cref{fig:geo_filter}, to adaptively express the geometry of a specific period,
we introduce a simple but effective geometry activation mechanism based on the predicted opacity.
During rendering, Gaussian primitives whose decoded opacity $\alpha_{ik}(t)$ falls below zero are excluded from both compositing and gradient backpropagation. 
This mechanism automatically deactivates anchors corresponding to geometry that is occluded in certain periods.
Consequently, the scaffold functions as a global geometric prior that dynamically activates relevant anchors per period,
preserving both adaptability and structural consistency.

\begin{figure}[!t]
    \centering
    \includegraphics[width=\linewidth]{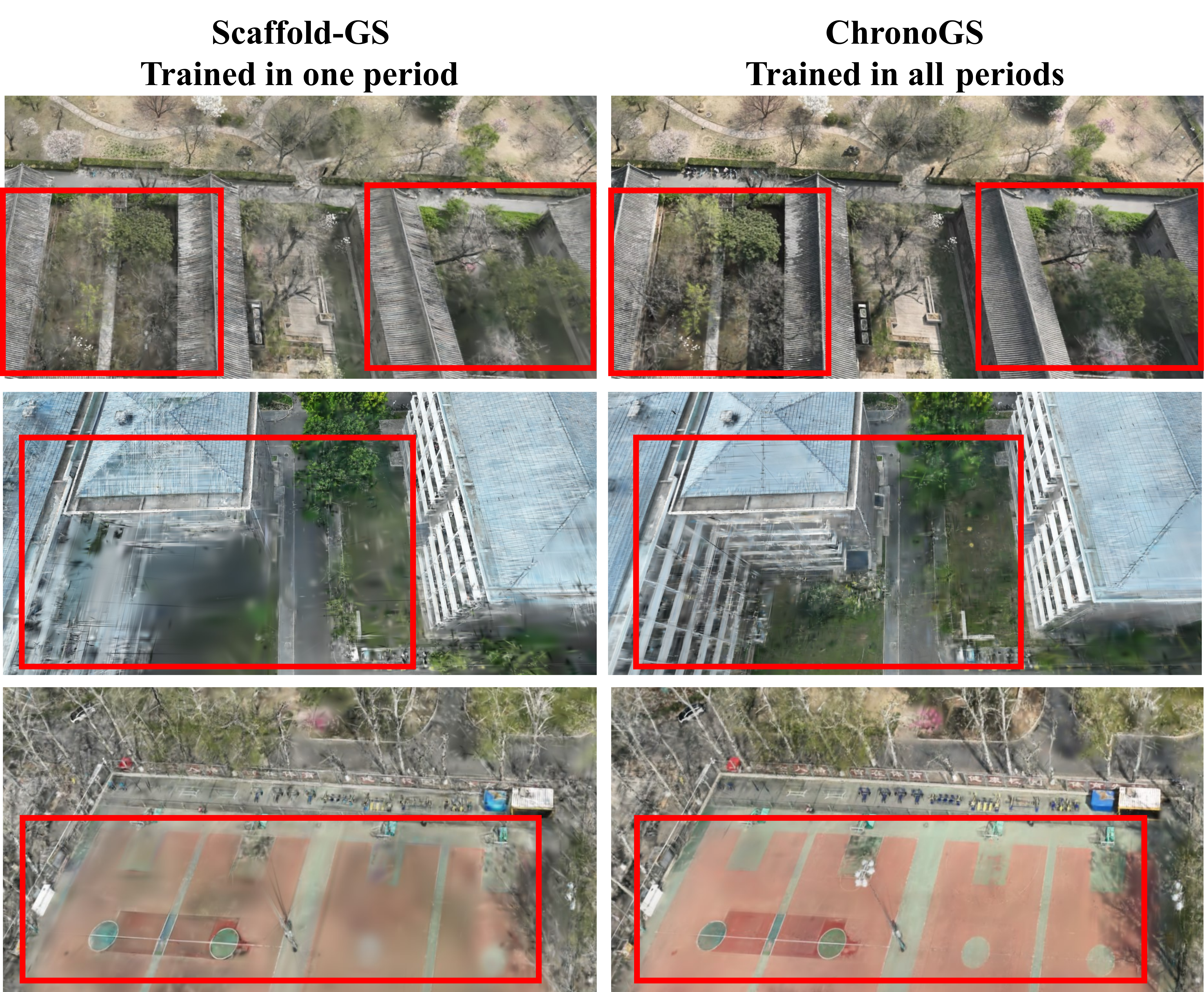}
    \vspace{-1.5em}
    \caption{Comparison between Scaffold-GS trained in one period (left) and ChronoGS trained across all periods (right).
Regions highlighted in red show that ChronoGS benefits from multi-period joint training, leveraging more observations to recover more accurate geometry and produce more complete reconstructions.}
\vspace{-1.5em}
    \label{fig:geo_vs_onetrain}
\end{figure}

\paragraph{Cross-period benefit.}
Training on multi-period imagery in a unified anchor scaffold also benefits geometrically stable regions. It allows complementary supervision on all periods' capture to improve reconstruction quality. 
Anchors that persist across periods receive denser supervision from multi-period observations,
leading to improved photometric consistency and reduced overfitting to single-period noise. This cross-period aggregation acts as an implicit regularizer: it encourages shared anchors to capture canonical geometry.
As shown in \cref{fig:geo_vs_onetrain}, such joint optimization produces sharper textures and a more reliable geometry in areas that remain static across periods.
Overall, the unified scaffold serves as a structural backbone that enables change geometry modeling. It also takes efficient use of all observations across periods in geometry-invariant areas, as opposed to training periods separately, which costs higher computation and storage.

\subsection{Gaussian Decoding and Rendering}
\label{subsec:decoder}

At each period $t$, we first fuse the features through temporal modulation:
\begin{equation}
\mathbf{h}_i(t) =
\mathbf{g}(t) \odot \mathbf{e}(t) + \mathbf{f}_i^{\text{var}}(t) \odot \mathbf{e}(t) +
\mathbf{f}_i^{\text{base}},
\end{equation}
where $\odot$ denotes channel-wise multiplication. 
Let $\mathbf{x}_i\!\in\!\mathbb{R}^3$ be the spatial position of anchor $a_i$, and 
$\mathbf{p}$ be the center of camera $C_j^t$. A lightweight MLP decoder $\Phi_{\theta}$ maps $\mathbf{h}_i(t)$ to the parameters of $K$ Gaussians:
\begin{equation}
\Big\{
\mathbf{S}_{ik}(t),\;
\alpha_{ik}(t),\;
\mathbf{c}_{ik}(t)
\Big\}_{k=1}^{K}
=
\Phi_{\theta}\!\big(\mathbf{h}_i(t),\,
\widehat{\mathbf{d}}_i\big),
\end{equation}
where $\mathbf{S}_{ik}(t)$ denotes the shape (rotation and scale),
$\alpha_{ik}(t)$ the opacity, $\mathbf{c}_{ik}(t)$ the color of the $k$-th Gaussian associated with anchor $a_i$, and $\widehat{\mathbf{d}}_i=\mathbf{x}_i-\mathbf{p}$ is the viewing direction from the camera to the anchor.
Each Gaussian’s 3D center is obtained as
$\boldsymbol{\mu}_{ik}=\mathbf{x}_i+\Delta\boldsymbol{\mu}_{ik}$,
where $\Delta\boldsymbol{\mu}_{ik}$ is a learnable offset
explicitly stored within anchor $a_i$.

Finally, the rendered color $\mathbf{C}(\mathbf{r},t)$ for a camera ray $\mathbf{r}$ follows the standard volumetric alpha compositing of 3DGS~\cite{kerbl20233d}. 
We first project 3D Gaussians to 2D with mean $\widehat{\boldsymbol{\mu}}_{ik}$ and covariance $\widehat{\boldsymbol{\Sigma}}_{ik}$.
Then, a front-to-back alpha compositing yields the rendered color:
\begin{equation}
\label{eq:compositing}
\widehat{\mathbf{C}}(\mathbf{u}) = 
\sum_{n} \widehat{\mathbf{c}}_{n}(\mathbf{u}) \widehat{\alpha}_{n}(\mathbf{u}) 
\prod_{j<n}(1-\widehat{\alpha}_{j}(\mathbf{u})).
\end{equation}

\subsection{Optimization}
\label{subsec:optimization}

We supervise the rendered images $\widehat{\mathbf{I}}$ with ground-truth observations $\mathbf{I}$ using a hybrid photometric objective:
\begin{equation}
\label{eq:recloss}
\mathcal{L} = 
\lambda \|\widehat{\mathbf{I}}-\mathbf{I}\|_1
+(1-\lambda) (1-\text{SSIM}(\widehat{\mathbf{I}},\mathbf{I})).
\end{equation}

Following \cite{lu2024scaffold}, we track the accumulated gradient magnitude of each anchor feature over recent iterations. 
Anchors with persistently small gradient magnitude are pruned to remove redundancy and reduce memory footprint. 
We also monitor the view-space positional gradients of the Gaussians emitted from each anchor.
When these gradients exceed a threshold, we spawn new anchors at the corresponding 3D locations obtained by the positions of Gaussians from old anchors. 
Both pruning and growth operations are triggered at regular training intervals,
allowing the scaffold to refine its structural support to cover the union of geometry across all periods
while maintaining a consistent canonical representation.


\section{Experiments}

\begin{figure}[]
    \centering
    \includegraphics[width=\linewidth]{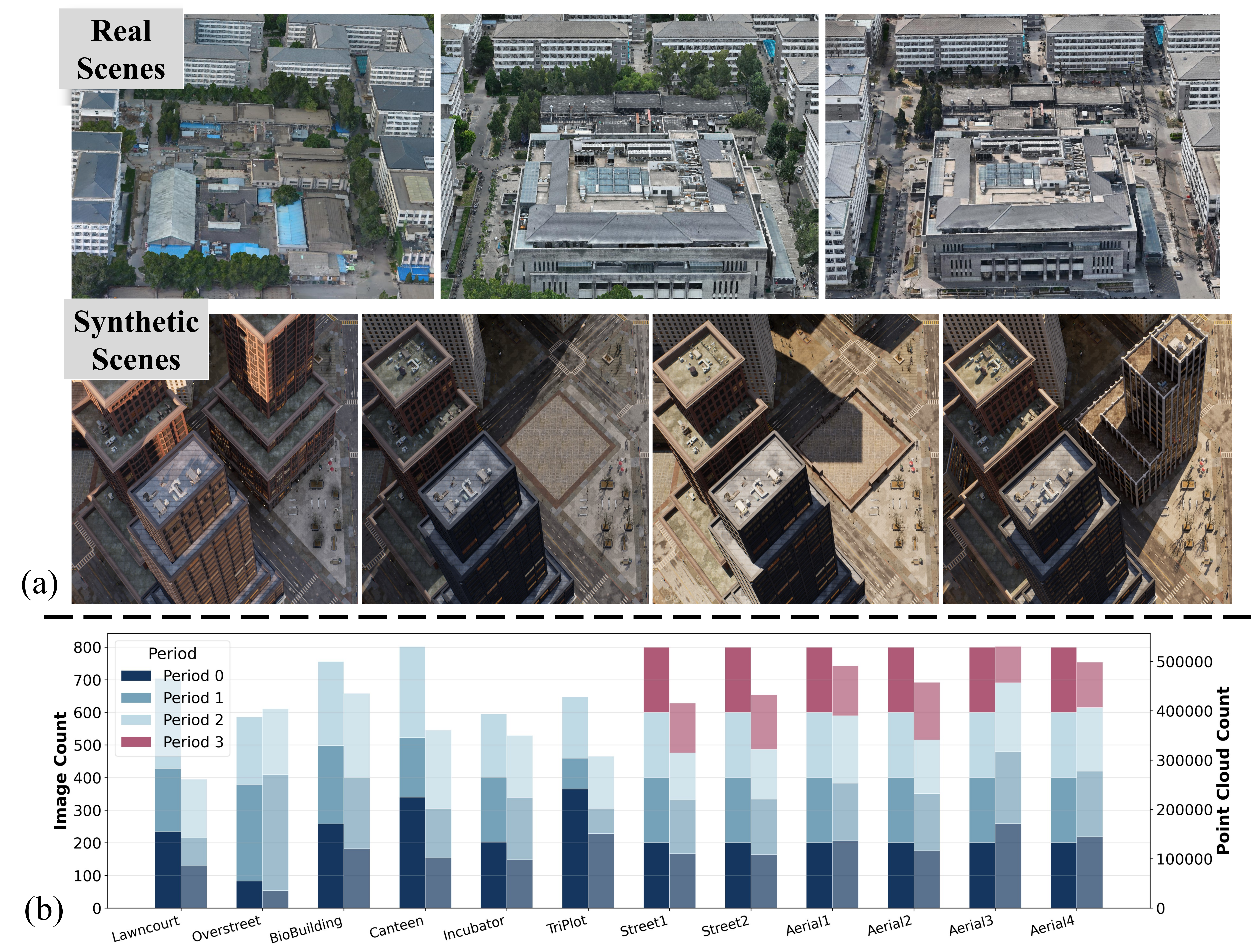}
    \vspace{-2em}

    \caption{\textbf{(a). Examples from the ChronoScene dataset.}
    Top: real-world aerial captures showing natural evolution across long temporal intervals (e.g., construction and season change).  
    Bottom: our synthetic scenes edited to simulate realistic changes with controllable geometry and appearance changes.
    \textbf{(b). Number distribution of image and point cloud across periods for 6 real-world and 6 synthetic scenes.} Left bars show image counts, right bars show sparse point cloud counts.
    }
    \vspace{-1.5em}
    \label{fig:dataset}
\end{figure}

\begin{figure*}
    \centering
    \includegraphics[width=\textwidth]{figs/result_new.pdf}
    \vspace{-2em}
    \caption{\textbf{Qualitative comparison on ChronoScene.}
      We compare ChronoGS with representative baselines. 
Each row shows novel-view renderings of different methods, along with ground truth. 
Static models trained on mixed multi-period data produce ghosting and appearance blending due to inconsistent geometry, 
while dynamic methods that assume continuous motion fail under the large temporal gaps, causing incorrect geometry interpolation. 
ChronoGS faithfully reconstructs period-specific geometry and appearance.
}

    \label{fig:qual_big}
\end{figure*}

\begin{figure*}[t]
    \centering
    \includegraphics[width=\textwidth]{figs/time.pdf}
    \vspace{-2.5em}
    \caption{\textbf{Temporal evolution at a test viewpoint.}
We query ChronoGS at continuous period indices $t\in[1,T]$ and render from the same camera.
Compared to dynamic baselines such as 4DGS that assume smooth motion, which often hallucinate intermediate structures or blend neighboring periods, 
ChronoGS cleanly switches between periods, recovering accurate geometry and appearance at each period. }
\vspace{-1em}
    \label{fig:temporal}
\end{figure*}

\begin{table*}[t]
  \centering
  \resizebox{\textwidth}{!}{%
  \begin{tabular}{@{}l|c|ccc|ccc|ccc|ccc|ccc|ccc@{}}
    \toprule
      \textbf{Real Scenes} & Avg. & \multicolumn{3}{c|}{Lawncourt} & \multicolumn{3}{c|}{Overstreet} & \multicolumn{3}{c|}{BioBuilding} & \multicolumn{3}{c|}{Canteen} & \multicolumn{3}{c|}{Incubator} & \multicolumn{3}{c}{TriPlot} \\
    \midrule
    Metrics  & Mem.$\downarrow$ & PSNR$\uparrow$ & SSIM$\uparrow$ & LPIPS$\downarrow$ & PSNR$\uparrow$  & SSIM$\uparrow$  & LPIPS$\downarrow$   & PSNR$\uparrow$ & SSIM$\uparrow$ & LPIPS$\downarrow$ & PSNR$\uparrow$ & SSIM$\uparrow$ & LPIPS$\downarrow$ & PSNR$\uparrow$ & SSIM$\uparrow$ & LPIPS$\downarrow$  & PSNR$\uparrow$  & SSIM$\uparrow$ & LPIPS$\downarrow$ \\
    \midrule
    3DGS\cite{kerbl20233d} & 1.01GB & 18.29 & 0.4658 & \cellcolor{orange!25}0.4862 & 18.79 & \cellcolor{yellow!25}0.6640 & \cellcolor{orange!25}0.3627 & 17.37 & \cellcolor{yellow!25}0.5754 & \cellcolor{orange!25}0.4402 & 19.64 & 0.5885 & \cellcolor{orange!25}0.4150 & 15.97 & 0.6232 & \cellcolor{yellow!25}0.3874 & 20.17 & \cellcolor{yellow!25}0.6320 & \cellcolor{orange!25}0.3793 \\
    Scaffold-GS\cite{lu2024scaffold} & 0.36GB & 16.27 & 0.3749 & 0.5363 & 15.20 & 0.5433 & 0.4390 & 16.09 & 0.4876 & 0.4923 & 17.70 & 0.5165 & 0.4528 & 15.72 & 0.5524 & 0.4291 & 18.04 & 0.5553 & 0.4280 \\
    GS-W\cite{zhang2024GS-W} & \cellcolor{yellow!25}0.23GB & \cellcolor{yellow!25}20.33 & 0.4018 & 0.5638 & \cellcolor{orange!25}21.64 & \cellcolor{orange!25}0.6644 & \cellcolor{yellow!25}0.3974 & \cellcolor{yellow!25}21.15 & 0.5536 & 0.5053 & 21.24 & 0.5227 & 0.4780 & \cellcolor{yellow!25}22.39 & \cellcolor{yellow!25}0.6788 & 0.3922 & \cellcolor{yellow!25}22.57 & 0.6064 & 0.4436 \\
    NeuSC\cite{lin2023neural}
    & 0.62GB & 14.40 & 0.1394 & 0.6871 & 12.48 & 0.0523 & 0.6825 & 13.82 & 0.0367 & 0.6707 & 16.56 & 0.3740 & 0.6613 & 12.81 & 0.0254 & 0.6859 & 17.27 & 0.4450 & 0.6195 \\
    CLNeRF\cite{cai2023clnerf}
    & \cellcolor{orange!25}0.19GB & 15.72 & 0.2148 & 0.7507 & 13.07 & 0.3851 & 0.7690 & 15.66 & 0.3501 & 0.7276 & 16.70 & 0.3168 & 0.7242 & 15.93 & 0.4333 & 0.7114 & 18.28 & 0.4089 & 0.6641 \\
    4DGS\cite{wu20244d} & \cellcolor{red!25}0.18GB & 19.03 & \cellcolor{yellow!25}0.4674 & 0.6085 & \cellcolor{yellow!25}19.12 & 0.5815 & 0.5110 & 17.18 & 0.5013 & 0.6211 & \cellcolor{yellow!25}21.67 & \cellcolor{yellow!25}0.5932 & 0.5345 & 18.39 & 0.5538 & 0.5333 & 19.35 & 0.5093 & 0.5691 \\
    realtime4DGS\cite{yang2023real} & 5.53GB & \cellcolor{orange!25}20.91 & \cellcolor{orange!25}0.5124 & \cellcolor{yellow!25}0.4943 & 17.17 & 0.5087 & 0.4993 & \cellcolor{orange!25}21.39 & \cellcolor{orange!25}0.5801 & \cellcolor{yellow!25}0.4920 & \cellcolor{orange!25}22.49 & \cellcolor{orange!25}0.6181 & \cellcolor{yellow!25}0.4292 & \cellcolor{orange!25}22.71 & \cellcolor{orange!25}0.7070 & \cellcolor{orange!25}0.3801 & \cellcolor{orange!25}22.86 & \cellcolor{orange!25}0.6681 & \cellcolor{yellow!25}0.3982 \\
    \hline
    Ours & 0.52GB & \cellcolor{red!25}22.16 & \cellcolor{red!25}0.6533 & \cellcolor{red!25}0.3390 & \cellcolor{red!25}22.66 & \cellcolor{red!25}0.7736 & \cellcolor{red!25}0.2838 & \cellcolor{red!25}22.87 & \cellcolor{red!25}0.7221 & \cellcolor{red!25}0.3274 & \cellcolor{red!25}25.15 & \cellcolor{red!25}0.7538 & \cellcolor{red!25}0.2647 & \cellcolor{red!25}23.79 & \cellcolor{red!25}0.7852 & \cellcolor{red!25}0.2721 & \cellcolor{red!25}24.16 & \cellcolor{red!25}0.7474 & \cellcolor{red!25}0.2901 \\
    \midrule
     \textbf{Synthetic Scenes} & Avg. &  \multicolumn{3}{c|}{Aerial1} & \multicolumn{3}{c|}{Aerial2} & \multicolumn{3}{c|}{Aerial3} & \multicolumn{3}{c|}{Aerial4} & \multicolumn{3}{c|}{Street1} & \multicolumn{3}{c}{Street2} \\
    \midrule
    Metrics  & Mem.$\downarrow$ &  PSNR$\uparrow$ & SSIM$\uparrow$ & LPIPS$\downarrow$ & PSNR$\uparrow$  & SSIM$\uparrow$  & LPIPS$\downarrow$   & PSNR$\uparrow$ & SSIM$\uparrow$ & LPIPS$\downarrow$ & PSNR$\uparrow$ & SSIM$\uparrow$ & LPIPS$\downarrow$ & PSNR$\uparrow$ & SSIM$\uparrow$ & LPIPS$\downarrow$  & PSNR$\uparrow$  & SSIM$\uparrow$ & LPIPS$\downarrow$ \\
    \midrule
    3DGS\cite{kerbl20233d} & 1.05GB & 20.79 & \cellcolor{orange!25}0.7831 & \cellcolor{orange!25}0.3179 & 19.95 & \cellcolor{orange!25}0.7877 & \cellcolor{orange!25}0.2252 & 23.20 & \cellcolor{orange!25}0.8250 & \cellcolor{orange!25}0.2682 & 21.41 & \cellcolor{orange!25}0.7934 & \cellcolor{orange!25}0.2631 & 15.40 & \cellcolor{yellow!25}0.6763 & \cellcolor{yellow!25}0.4065 & 18.77 & \cellcolor{yellow!25}0.7225 & \cellcolor{yellow!25}0.3580 \\
    Scaffold-GS\cite{lu2024scaffold} & 0.30GB & 19.52 & 0.7273 & 0.3605 & 18.35 & \cellcolor{yellow!25}0.7199 & 0.2738 & 19.51 & 0.7485 & 0.3327 & 18.38 & 0.7020 & 0.3272 & 14.49 & 0.5905 & 0.5252 & 16.86 & 0.5856 & 0.5279 \\
    GS-W \cite{zhang2024GS-W} & \cellcolor{yellow!25}0.22GB & \cellcolor{orange!25}25.11 & 0.7581 & 0.3754 & \cellcolor{orange!25}23.21 & 0.7005 & \cellcolor{yellow!25}0.2510 & \cellcolor{orange!25}28.79 & 0.7958 & 0.3186 & \cellcolor{orange!25}26.13 & \cellcolor{yellow!25}0.7826 & \cellcolor{yellow!25}0.2644 & 15.73 & 0.6328 & 0.4847 & 18.93 & 0.7075 & 0.3873 \\
    NeuSC\cite{lin2023neural}
    & 0.62GB & 15.88 & 0.1929 & 0.6244 & 14.85 & 0.2385 & 0.6512 & 16.82 & 0.1996 & 0.5843 & 16.69 & 0.3160 & 0.6230 & 11.60 & 0.0778 & 0.6256 & 14.53 & 0.1955 & 0.6630 \\
CLNeRF\cite{cai2023clnerf}
    & \cellcolor{orange!25}0.20GB & 20.47 & 0.5892 & 0.6513 & \cellcolor{yellow!25}22.21 & 0.5254 & 0.5950 & \cellcolor{yellow!25}26.07 & 0.6850 & 0.5148 & \cellcolor{yellow!25}23.89 & 0.6045 & 0.5699 & \cellcolor{orange!25}21.25 & 0.5866 & 0.6046 & \cellcolor{orange!25}23.17 & 0.6169 & 0.5833 \\
    4DGS\cite{wu20244d} & \cellcolor{red!25}0.18GB  & 20.69 & 0.7140 & 0.4375 & 19.62 & 0.6325 & 0.4324 & 22.47 & 0.7359 & 0.3971 & 20.67 & 0.6675 & 0.4369&  13.53 & 0.5440 & 0.5635 & 16.03 & 0.5310 & 0.5928 \\
    realtime4DGS\cite{yang2023real} & 7.41GB & \cellcolor{yellow!25}22.29 & \cellcolor{yellow!25}0.7774 & \cellcolor{yellow!25}0.3351 & 20.04 & 0.7088 & 0.3268 & 23.93 & \cellcolor{yellow!25}0.8044 & \cellcolor{yellow!25}0.3075 & 21.60 & 0.7443 & 0.3341 & \cellcolor{yellow!25}18.88 & \cellcolor{orange!25}0.7121 & \cellcolor{orange!25}0.3944 & \cellcolor{yellow!25}20.53 & \cellcolor{orange!25}0.7451 & \cellcolor{orange!25}0.3514  \\
    \hline
    Ours & 0.65GB & \cellcolor{red!25}28.80 & \cellcolor{red!25}0.8562 & \cellcolor{red!25}0.2509 & \cellcolor{red!25}26.50 & \cellcolor{red!25}0.8467 & \cellcolor{red!25}0.1854 & \cellcolor{red!25}30.11 & \cellcolor{red!25}0.8751 & \cellcolor{red!25}0.2266 & \cellcolor{red!25}28.10 & \cellcolor{red!25}0.8350 & \cellcolor{red!25}0.2305 & \cellcolor{red!25}24.70 & \cellcolor{red!25}0.8071 & \cellcolor{red!25}0.3241 & \cellcolor{red!25}26.56 & \cellcolor{red!25}0.8220 & \cellcolor{red!25}0.2946 \\
    \bottomrule
  \end{tabular}
  }
  \vspace{-.5em}
  \caption{
Quantitative comparison on ChronoScene dataset.
The best, second-best, and third-best results in each column are highlighted by 
\colorbox{red!25}{red}, \colorbox{orange!25}{orange}, and \colorbox{yellow!25}{yellow}, respectively.
The upper part shows results on real scenes, and the lower part on synthetic scenes.
}
\vspace{-1em}
  \label{tab:result_combined}
\end{table*}

\subsection{Implementation Details}

We set anchor's base feature dimension to $d_b=16$, the local period-varying feature dimension to $d_v=16$, and the global feature dimension to $d_g=32$.
For each anchor, the lightweight MLP decoder predicts $K=10$ Gaussians.
All experiments are trained for 40k iterations to ensure sufficient convergence on large-scale multi-period scenes.
Details and ablations on hyperparameters and training strategy are provided in the supplementary material.

\subsection{ChronoScene Dataset}

We introduce \emph{ChronoScene}, a multi-period dataset designed to challenge long-term 3D reconstruction under both \emph{geometric} and \emph{appearance} evolution. As illustrated in \cref{fig:dataset}, scenes in ChronoScene exhibit coupled changes such as building construction or demolition, vegetation growth, and variations in season, lighting, and weather. ChronoScene contains a total of \textbf{8,891 images} across \textbf{12 scenes} and \textbf{42 sub-scenes} (each sub-scene corresponds to one specific scene at one period). All images are registered into a \emph{common world coordinate frame} in COLMAP~\cite{schoenberger2016sfm} format, ensuring that cameras across periods are spatially comparable.

\noindent\textbf{Real data.}  
The real-world scenes span time gaps ranging from one year to three years between periods, reflecting natural changes in urban environments. Images are geo-tagged at capture time and globally aligned using GPS priors, followed by human-in-the-loop refinement to ensure consistent cross-period camera registration.

\noindent\textbf{Synthetic data.}  
To complement real data with controlled, precisely known temporal changes, we build synthetic scenes based on Matrix City~\cite{li2023matrixcity}. We manually edit urban assets to simulate realistic evolutions—such as new constructions, façade refurbishments, or vegetation changes—and render both aerial and street-level views.  

We release all RGB images, calibrated camera poses, and per-image period identifiers. More statistics, licensing, and details are provided in the supplementary material.

\subsection{Evaluation}

Unless otherwise noted, experiments in the main paper are conducted on \textbf{ChronoScene}, our primary benchmark for multi-period geometric and appearance evolution.
To further assess generalization, we additionally report quantitative results on WAT\cite{cai2023clnerf}, CL-Splat\cite{ackermann2025cl}, and NeuSC\cite{lin2023neural} datasets in the supplementary material.

\noindent \textbf{Baselines setup.}
We compare representative paradigms covering diverse settings: 3DGS~\cite{kerbl20233d} as the canonical static baseline, Scaffold-GS~\cite{lu2024scaffold} introducing anchor-based structured Gaussians, GS-W~\cite{zhang2024GS-W} which targets uncontrolled photo collections, NeuSC~\cite{lin2023neural} for long-term in-the-wild appearance changes, CLNeRF~\cite{cai2023clnerf} for incremental scene updates, as well as 4DGS~\cite{wu20244d} and Realtime4DGS~\cite{yang2023real} that extend Gaussians to temporally continuous motion.

\noindent \textbf{Protocol.}
Camera trajectories are independently captured at different periods, so we assess reconstruction quality \emph{within each period}: for each test set of a given period, we render from its held-out cameras and compute PSNR, SSIM, and LPIPS with ground-truth images; metrics are then averaged over periods. We ensure that all methods use the same train-test splits and data augmentations.

\begin{table}[t]
  \centering
  \tabcolsep=0.1cm
  \resizebox{\linewidth}{!}{
  \begin{tabular}{@{}l|ccc|ccc|ccc@{}}
    \toprule
      & \multicolumn{3}{c|}{3DGS~\cite{kerbl20233d}} & \multicolumn{3}{c|}{Scaffold-GS~\cite{lu2024scaffold}} & \multicolumn{3}{c}{Ours} \\
    \midrule
    Period & PSNR$\uparrow$ & Mem.$\downarrow$ & Iters. 
           & PSNR$\uparrow$ & Mem.$\downarrow$ & Iters. 
           & PSNR$\uparrow$ & Mem.$\downarrow$ & Iters. \\
    \midrule
    Period 0 & 21.97 & 1.6GB  & 40k 
             & 22.04 & 0.19GB & 40k 
             & \textbf{22.64} & N/A      & N/A \\
    Period 1 & 22.31 & 2.0GB  & 40k 
             & 21.88 & 0.25GB & 40k 
             & \textbf{22.50} & N/A     & N/A \\
    Period 2 & 22.23 & 1.3GB & 40k 
             & 21.81 & 0.21GB   & 40k 
             & \textbf{22.88} & N/A      & N/A \\
    Overall  & 22.23 & 4.9GB  & 120k 
             & 21.87 & 0.65GB & 120k 
             & \textbf{22.66} & \textbf{0.57GB} & \textbf{40k} \\
    \bottomrule
  \end{tabular}
  }
  \vspace{-.5em}
  \caption{
Quantitative comparison of static methods' per-period independent training and our joint multi-period optimization on the \textit{Overstreet} scene.
Reported metrics include PSNR, memory footprint, and training iterations.
ChronoGS achieves the \textbf{highest} quality across all periods while markedly reducing training iterations and storage requirements compared with static baselines.}
  \label{tab:joint_vs_sep}
\end{table}

\subsection{Results}
\label{subsec:results}

\noindent\textbf{Quantitative performance.}
We report results separately for the real and synthetic portions of ChronoScene. \cref{tab:result_combined} summarizes image-level metrics and overall storage cost in all 12 scenes of real and synthetic data, where geometry and appearance changes across periods. Our method consistently outperforms existing baselines on ChronoScene, achieving the best metrics across scenes while maintaining a compact memory footprint.
These results confirm the effectiveness of our method in modeling both geometry and appearance variations across periods.

\noindent\textbf{Qualitative comparisons.}
\cref{fig:qual_big} presents a side-by-side comparison of our renderings, ground truth, and representative baselines for held-out views in each period. Static models tend to blur conflicting configurations and dynamic models produce temporal smearing; in contrast, our approach maintains sharp, period-correct details. Our method preserves period-specific appearance and suppresses cross-period ghosting that appears in static models. 

\noindent\textbf{Temporal interpolation.}
To evaluate whether our model can synthesize intermediate periods, in \cref{fig:temporal} we show the renderings at the \emph{same} test viewpoint across periods and compare our method to dynamic baselines.
Methods designed for continuous dynamics often interpolate through non-existent intermediate states under ChronoScene’s discontinuous region, leading to hallucinated transitions or residuals from neighboring periods.
Our model cleanly switches between periods with more reasonable intermediate states.

\noindent\textbf{Compare with Per-Period Training.} 
To evaluate the effect of multi-period joint optimization, we compare ChronoGS with static baselines trained independently for each period under the same evaluation setting on \textit{Overstreet} scene.
As summarized in \cref{tab:joint_vs_sep} and shown in \cref{fig:geo_vs_onetrain}, compared with training separately in each period's capture, our approach, which trains data from all periods together, receives denser and more complementary supervision—leading to the same or improved quality and significant reductions in both training and storage cost.

\begin{table}[]
  \centering
  \tabcolsep=0.1cm
  \resizebox{.9\linewidth}{!}{%
 \begin{tabular}{@{}l|ccc|ccc@{}}
    \toprule
      & \multicolumn{3}{c|}{Lawncourt} & \multicolumn{3}{c}{Canteen} \\
    \midrule
    Settings  & PSNR$\uparrow$ & SSIM$\uparrow$ & LPIPS$\downarrow$  & PSNR$\uparrow$ & SSIM$\uparrow$ & LPIPS$\downarrow$ \\
    \midrule
    w/o Var.\&Global. & 18.18 & 0.4609 & 0.4928 & 19.88 & 0.5959 & 0.4107 \\
    w/o Base.  & 21.60 & 0.6055 & 0.3918 & 24.57 &0.7289 &0.2970 \\
    w/o Var. & 21.96 & 0.6440 & 0.3526& 24.62 & 0.7426 & 0.2801  \\
    w/o Global. & 22.11 & 0.6466 & 0.3497& 24.94 & 0.7502 & 0.2716  \\
    \textbf{Ours (full)} &  \textbf{22.16} & \textbf{0.6533} & \textbf{0.3390} & \textbf{25.15} & \textbf{0.7538} & \textbf{0.2647} \\
    \bottomrule
  \end{tabular}
  }
  \vspace{-.5em}
  \caption{
  Quantitative ablation results on two real scenes \textit{Lawncourt} and \textit{Canteen}.
  Removing the base, local period-varying, or global period-varying feature leads to performance degradation,
  confirming that all three are essential in our design.}
  \vspace{-1.5em}
  \label{tab:ablation}
\end{table}

\begin{figure}[!t]
    \centering
    \includegraphics[width=.98\linewidth]{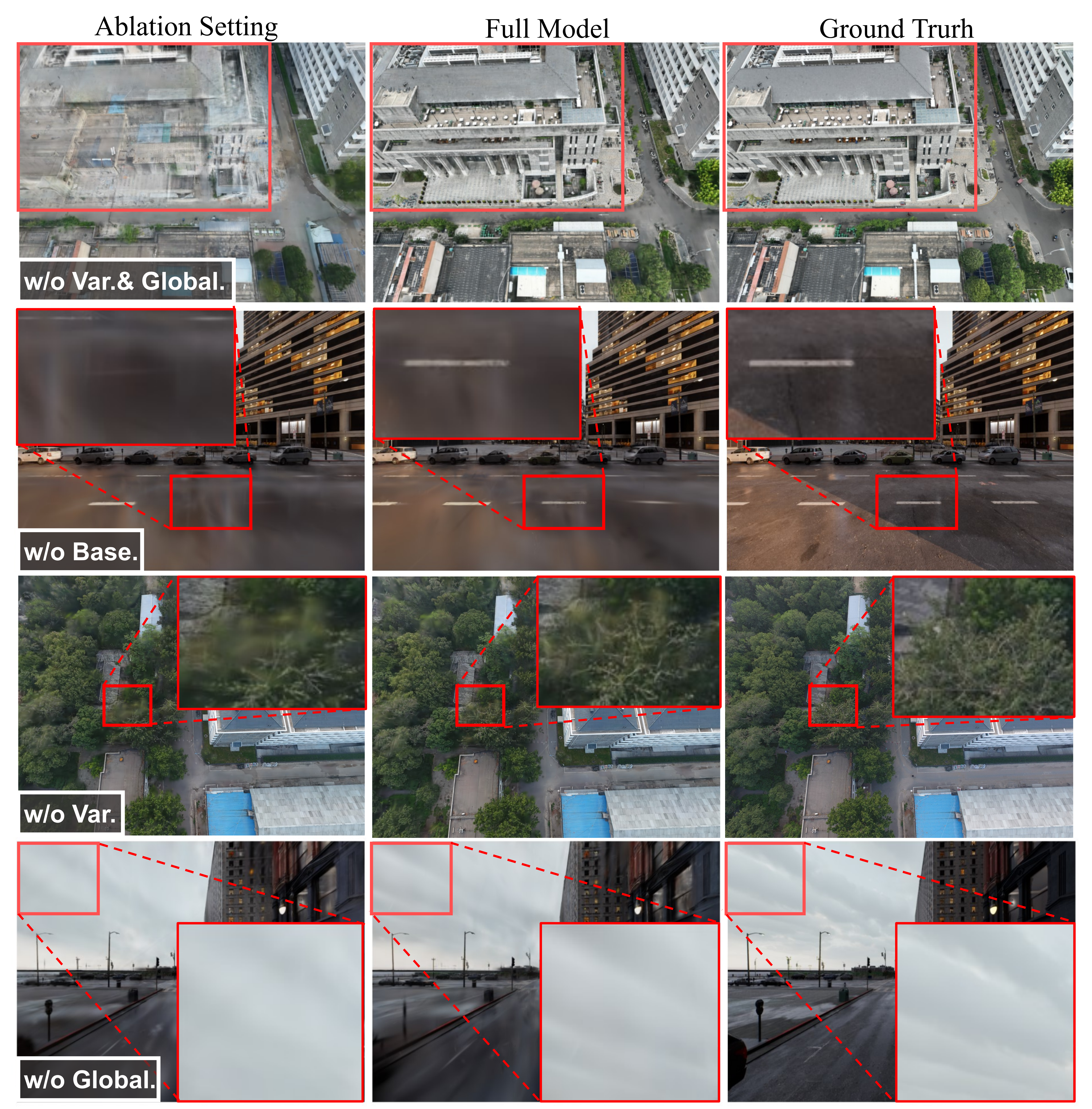}
    \vspace{-.5em}
    \caption{
\textbf{Qualitative Ablation results.} Removing the base, local period-varying, or global period-varying feature leads to degraded geometry stability, local temporal fidelity, and scene-level appearance consistency, respectively, compared with the full model.
    }
    \vspace{-1.5em}
    \label{fig:ablation}
\end{figure}

\subsection{Ablation Study}
\label{subsec:ablation}

To evaluate the role of each component in ChronoGS, we perform ablations under identical training settings.
We compare the full model with several degraded variants: 
\textbf{w/o Base.}, removing the base feature $\mathbf{f}^{\text{base}}_i$ and forcing each anchor to rely only on period-dependent representations;
\textbf{w/o Var.}, removing the local period-varying feature $\mathbf{f}^{\text{var}}_i(t)$ and disabling local temporal adaptation;
\textbf{w/o Global.}, removing the global period-varying feature $\mathbf{g}(t)$ thus losing global appearance coherence; and \textbf{w/o Var.\&Global.}, removing $\mathbf{g}(t)$ and $\mathbf{f}^{\text{var}}_i(t)$ to disable model's temporal behavior.

As shown in \cref{tab:ablation}, the full ChronoGS achieves the best quantitative and perceptual scores. 
Our design disentangles changes from period-invariant structure to achieve consistent multi-period scene reconstruction. Without the base feature, the shared geometry becomes unstable. Without the local period-varying feature, local temporal changes cannot be expressed, and without the global period-varying feature, scene-level illumination and color consistency degrade noticeably. These can be clearly observed in \cref{fig:ablation}.

\section{Conclusion}


We presented \textbf{ChronoGS}, a temporally modulated Gaussian representation for reconstructing multi-period scenes.
ChronoGS represents a multi-period scene with a unified anchor scaffold that aggregates geometry shared across periods. Temporally modulated features and a geometry activation mechanism jointly model both invariant structures and period-varying attributes.
This formulation effectively handles geometric and appearance variations that arise when imagery is collected in different periods.
In parallel, we introduced \textbf{ChronoScene}, a novel benchmark that collects multi-period observations and captures both geometric and appearance evolution across periods.
Extensive experiments demonstrate that ChronoGS achieves sharper renderings, more accurate geometry, and improved cross-period consistency compared to existing baselines. Ablation studies validate the contribution of each component.
To our knowledge, this is the first work to reconstruct multi-period scenes within a unified, differentiable Gaussian-splatting framework that jointly handles non-continuous geometry and appearance changes. We provide both a simple but strong baseline and a valuable public benchmark for future research.



\section*{Acknowledgments}
This work is supported by the National Key R\&D Program of China (Grant No. 2023YFB3309001).
This work is also supported by the Shenzhen Science and Technology Program (Grant No. KJZD20230923114114028) and the Shandong Key Research and Development Program (Grant No. 2024CXGC010801).

{
    \small
    \putbib[main]
}

\end{bibunit}

\appendix
\begin{bibunit}

\clearpage
\setcounter{page}{1}
\maketitlesupplementary

\section{More Implementation  Details}

To help readers better understand our method and reproduce our results, we
provide additional implementation details in this section.
For clarity and completeness, we only summarize the most relevant aspects here.
Please refer to our code for further details and full implementation.

\subsection{Environment Setup}

All experiments are conducted on a Linux workstation equipped with NVIDIA RTX A6000 48GB
GPUs and Intel Xeon Gold 6330 (112) @ 3.1 GHz cpus. We use PyTorch 2.4.1 with CUDA 12.1 and Python 3.8.19. We use the gsplat\cite{ye2025gsplat} backend and differentiable splatting kernels provided therein. Training and
evaluation on large images are performed with the default downsampling strategy of 3DGS\cite{kerbl20233d}.

\subsection{Anchor Scaffold Construction}
\label{sec:scaffold}

We follow the anchor-based scene parameterization introduced in Scaffold-GS\cite{lu2024scaffold}.
Given a COLMAP\cite{schoenberger2016sfm} sparse reconstruction, let 
$\mathcal{P}=\{\mathbf{p}_j\}$ denote the sparse SfM point cloud.
The goal is to construct a compact yet spatially uniform set of 
\emph{anchors} $\mathcal{A}=\{a_i\}_{i=1}^{N_a}$ that provide stable
geometric support for subsequent Gaussian decoding.

\paragraph{Spatial Grid Generation.}
The SfM point cloud is first enclosed by an axis-aligned bounding box.
We discretize the bounding box into a 3D rectilinear grid with cell size
$\Delta=0.001$ unless otherwise specified.
For each grid cell $\mathcal{G}_u$, we check whether it contains at least
one SfM point:
\[
\mathcal{G}_u \text{ is occupied} \iff 
\exists \, \mathbf{p}_j \in \mathcal{P} 
\text{ s.t. } \mathbf{p}_j \in \mathcal{G}_u.
\]

\paragraph{Anchor Feature Initialization.}
Each anchor $a_i$ maintains three types of learnable feature vectors:
a period-invariant base feature $f_i^{\text{base}} \in \mathbb{R}^{d_b}$,
a set of local period-varying features 
$f_i^{\text{var}}(t) \in \mathbb{R}^{T \times d_v}$.
And a global period feature $g(t) \in \mathbb{R}^{T \times d_g}$ shared across anchors.
Unless otherwise specified, all anchors' feature vectors are initialized with zero
initialization, which allows the model
to learn meaningful structure purely from data. At initialization, the anchor
distribution mainly captures the coarse occupied volume of the scene, while the
detailed geometry and appearance are progressively refined through the Gaussian
clusters decoded from these features during training.

\paragraph{Anchor Filter.}
To accelerate rendering, we precompute a coarse visibility mask for each anchor.
For every camera $C_j$, an anchor $a_i$ is marked as potentially
visible if it lies inside the camera frustum and in front of the camera center.
This frustum-based filtering significantly reduces runtime culling overhead and
is reused by our method without modification.

\paragraph{Anchor Growth and Pruning.}
We maintains a dynamic anchor set instead of a fixed scaffold. During
training, anchors are periodically added or removed based on simple
reconstruction statistics. In the growth step, regions with large residuals or
strong gradients are identified in a coarse-to-fine hierarchy, and new anchors
are inserted to increase local capacity. In the pruning step, anchors with
consistently low visibility, opacity, or gradient magnitude are removed as
redundant. This anchor-level adaptation keeps the scaffold compact while
allocating more anchors to challenging regions.

\subsection{Gaussian Cluster Prediction}
\label{sec:cluster_prediction}

Each anchor decodes a local cluster of Gaussian primitives,
which collectively represent local shape and appearance. Instead of using a single large MLP to predict all Gaussian parameters, we use
three lightweight MLPs to decode geometry, opacity, and appearance separately.
We find this design more stable in practice.
We use $K=10$ Gaussians per anchor. Our MLPs employ a lightweight two-layer architecture: a fully connected layer with ReLU activation mapping the input to a hidden representation of dimension $d_f$, followed by an output layer producing task-specific parameters. The opacity MLP uses $\tanh$ activation, the color MLP uses $\text{sigmoid}$ activation, and the covariance MLP has no final activation to allow unconstrained geometric parameters.

\subsection{Training Strategy}

We train our model using the Adam optimizer and employing differentiated learning rates for different parameter groups. Anchor positions remain fixed throughout training, while offset parameters use an exponential decay schedule from $0.01 \times \text{spatial\_lr\_scale}$ to $0.0001 \times \text{spatial\_lr\_scale}$, where spatial\_lr\_scale corresponds to the scale of scene. MLP networks for opacity, covariance, and color prediction use exponential decay schedules with initial learning rates of $0.002$, $0.004$, and $0.008$, decaying to $0.00002$, $0.004$, and $0.00005$, respectively. Geometric parameters, including anchor features, scaling, and rotation use fixed learning rates of $0.0075$, $0.007$, and $0.002$. Our training process consists of three distinct phases: during the initialization phase (iterations 0--500), we initialize model parameters and begin standard gradient-based optimization; in the densification phase (iterations 500--20,000), we first collect gradient and opacity statistics from iterations 500--1,500, then perform active densification every 100 iterations from iteration 1,500 onwards, where we grow new anchors at offsets with gradient norms exceeding the threshold $\tau_g = 0.0002$ and visibility above $0.5 \times \text{success\_threshold} \times \text{update\_interval}$, while pruning anchors with accumulated opacity below $\text{min\_opacity} \times \text{anchor\_demon}$ and sufficient training statistics; finally, in the pure optimization phase (iterations 20,000--40,000), we disable densification, clean up accumulated statistics, and continue pure parameter optimization without structural changes to the anchor set.

\section{More Ablation Studies}

\subsection{Ablation on Feature Dimensions}

\begin{table}
  \centering
  \tabcolsep=0.1cm
    \resizebox{.85\linewidth}{!}{%
 \begin{tabular}{@{}l|cccc@{}}
    \toprule
      & \multicolumn{4}{c}{Scene Street1}  \\
    \midrule
    Settings  & PSNR$\uparrow$ & SSIM$\uparrow$ & LPIPS$\downarrow$ & Mem.$\downarrow$ \\
    \midrule
    $\text{dim}(g(t))=16 * T$. & 24.55 & \textbf{0.7991} & 0.3333 & 0.42GB \\
    $\text{dim}(g(t))=24 * T$. & 24.25 & 0.7948 & 0.3389 &  \textbf{0.36GB} \\
    $\text{dim}(g(t))=\textbf{32} * T$.& \textbf{24.57} & 0.7972 &\textbf{0.3241}  & 0.52GB  \\
    $\text{dim}(g(t))=48 * T$.& 23.94 & 0.7790 & 0.3623 & 0.58GB  \\
    $\text{dim}(g(t))=64 * T$.& 24.25 & 0.7921 & 0.3426 & 0.43GB \\

    \bottomrule
  \end{tabular}
  }
  \caption{\textbf{Ablation on the dimension $T \times d_g$ of the global feature $g(t)$.}
We evaluate on Scene \textit{Street1}. 
A small dimension limits scene-level change modeling, while a large one increases
memory with no consistent quality gain. 
The best balance is achieved at $d_g=32$.}
  \label{tab:ablation_g}
\end{table}

\begin{table}
  \centering
  \tabcolsep=0.1cm
    \resizebox{.9\linewidth}{!}{%
 \begin{tabular}{@{}l|cccc@{}}
    \toprule
      & \multicolumn{4}{c}{Scene Aerial1}  \\
    \midrule
    Settings  & PSNR$\uparrow$ & SSIM$\uparrow$ & LPIPS$\downarrow$ & Mem.$\downarrow$ \\
    \midrule
    $\text{dim}(f_i^{\text{var}}(t))=8*T$. &28.35 & 0.8478 & 0.2631 & \textbf{0.49GB} \\

    $\text{dim}(f_i^{\text{var}}(t))=\textbf{16}*T$.& \textbf{28.80} & \textbf{0.8562} & \textbf{0.2509} & 0.63GB  \\
    $\text{dim}(f_i^{\text{var}}(t))=24*T$.& 28.62 & 0.8521 & 0.2573 & 0.90GB  \\
    $\text{dim}(f_i^{\text{var}}(t))=32*T$.&28.61 & 0.8512 & 0.2572 & 0.74GB \\
    \bottomrule
  \end{tabular}
  }
  \caption{\textbf{Ablation on the dimension $T\times d_v$ of the global feature $f_i^\text{var}(t)$.}
We evaluate on Scene \textit{Aerial1}. 
A small dimension limits anchor-level structure modeling, while a large one increases
memory with no consistent quality gain. 
So we choose $d_v=16*T$.}
  \label{tab:ablation_v}
\end{table}

\begin{table}
  \centering
  \tabcolsep=0.1cm
    \resizebox{.85\linewidth}{!}{%
 \begin{tabular}{@{}l|cccc@{}}
    \toprule
      & \multicolumn{4}{c}{Scene Aerial2}  \\
    \midrule
    Settings  & PSNR$\uparrow$ & SSIM$\uparrow$ & LPIPS$\downarrow$ & Mem.$\downarrow$ \\
    \midrule
    $\text{dim}(f_i^{\text{base}})=8$. & 26.00 & 0.8340 & 0.2011 & \textbf{0.87GB} \\
    $\text{dim}(f_i^{\text{base}})=\textbf{16}$.& \textbf{26.50} & 0.8467 & 0.1854 & 1.10GB  \\
    $\text{dim}(f_i^{\text{base}})=24$. & 26.41 & \textbf{0.8478} & \textbf{0.1850} & 1.20GB   \\
    $\text{dim}(f_i^{\text{base}})=32$.& 26.38 & 0.8452 & 0.1867 & 1.30GB \\

    \bottomrule
  \end{tabular}
  }
  \caption{\textbf{Ablation on the dimension $d_b$ of the global feature $f_i^\text{base}$.}
We evaluate on Scene \textit{Aerial2}. 
A small dimension limits scene-level appearance modeling, while a large one exhibits no dominant quality gain but costs more storage. 
We use $d_b=16$ for the trade-off.}
  \label{tab:ablation_b}
\end{table}

We further ablate the dimensions of the three feature components in our
period-aware representation: the global period-varying feature $g(t)$, the local period-varying
feature $f_i^{\text{var}}(t)$, and the base feature $f_i^{\text{base}}$. These
dimensions control the capacity allocated to scene-level period-varying changes,
local period-specific variations, and stable, period-invariant structure.

For $g(t)$, as shown in \cref{tab:ablation_g}, we sweep $d_g \in \{16,24,32,48,64\}$ and observe that
$d_g=32$ offers the best quality. For the
period-varying feature $f_i^{\text{var}}(t)$, as shown in \cref{tab:ablation_v}, we test $d_v \in \{8,16,24,32\}$,
where small $d$ restricts temporal expressiveness, and large $d$ increases cost
with diminishing returns. For the base feature $f_i^{\text{base}}$, as shown in \cref{tab:ablation_b}, we evaluate
$d_b \in \{8,16,24,32\}$ and find that moderate dimensions (e.g., $d_b=16$)
provide the best trade-off between accuracy and memory cost.

Overall, the configuration $(d_b, d_v, d_g) = (16, 16, 32)$ provides the best
balance between reconstruction quality and storage cost.

\subsection{Effect of Anchor Feature Initialization}

\begin{table}
  \centering
  \tabcolsep=0.1cm
    \resizebox{.6\linewidth}{!}{%
 \begin{tabular}{@{}l|ccc@{}}
    \toprule
      & \multicolumn{3}{c}{Scene Street2}  \\
    \midrule
    Settings  & PSNR$\uparrow$ & SSIM$\uparrow$ & LPIPS$\downarrow$ \\
    \midrule
    $\text{Uni. Init.}$ & 26.47 & 0.8182 & 0.3011  \\
    $\text{Gaussian Init.}$ & 26.11 & 0.8155 & 0.3024    \\
    $\text{Zero Init.}$ & \textbf{26.56} & \textbf{0.8220} & \textbf{0.2946}   \\
    \bottomrule
  \end{tabular}
  }
  \caption{\textbf{Ablation on anchor feature initialization.} 
Evaluated on \textit{Street2}. 
We compare three initialization strategies for anchor features: uniform sampling,
Gaussian sampling, and zero initialization. Zero initialization yields the best
overall reconstruction quality, suggesting that our framework does not require
pre-imposed feature diversity to achieve stable optimization.}
  \label{tab:ablation_init}
\end{table}

\begin{table*}[!htbp]
\centering
\resizebox{\textwidth}{!}{
\begin{tabular}{lcccccccccc}
\toprule
Scene & \multicolumn{2}{c}{Period 0} & \multicolumn{2}{c}{Period 1} & \multicolumn{2}{c}{Period 2} & \multicolumn{2}{c}{Period 3} & \multicolumn{2}{c}{Total} \\
 & Images & Points & Images & Points & Images & Points & Images & Points & Images & Points \\
\midrule
Lawncourt & 234 & 85,655 & 193 & 57,884 & 277 & 117,595 & N/A & N/A & 704 & 261,134 \\
Overstreet & 83 & 35,513 & 295 & 235,537 & 208 & 132,957 & N/A & N/A & 586 & 404,007 \\
BioBuilding & 258 & 120,122 & 240 & 143,694 & 258 & 171,462 & N/A & N/A & 756 & 435,278 \\
Canteen & 340 & 101,523 & 183 & 99,915 & 279 & 159,250 & N/A & N/A & 802 & 360,688 \\
Incubator & 202 & 98,073 & 199 & 126,315 & 194 & 125,461 & N/A & N/A & 595 & 349,849 \\
TriPlot & 365 & 150,896 & 94 & 50,081 & 189 & 106,937 & N/A & N/A & 648 & 307,914 \\
Street1 & 200 & 110,591 & 200 & 108,932 & 200 & 95,038 & 200 & 101,074 & 800 & 415,635 \\
Street2 & 200 & 108,621 & 200 & 112,293 & 200 & 101,061 & 200 & 110,714 & 800 & 432,689 \\
Aerial1 & 200 & 136,491 & 200 & 116,790 & 200 & 136,624 & 200 & 101,494 & 800 & 491,399 \\
Aerial2 & 200 & 116,443 & 200 & 115,481 & 200 & 109,213 & 200 & 116,434 & 800 & 457,571 \\
Aerial3 & 200 & 171,323 & 200 & 145,781 & 200 & 139,733 & 200 & 73,597 & 800 & 530,434 \\
Aerial4 & 200 & 144,446 & 200 & 133,058 & 200 & 129,169 & 200 & 92,044 & 800 & 498,717 \\
\midrule
\textbf{Total} & \textbf{2,682} & \textbf{1,379,697} & \textbf{2,404} & \textbf{1,445,761} & \textbf{2,605} & \textbf{1,524,500} & \textbf{1,200} & \textbf{595,357} & \textbf{8,891} & \textbf{4,945,315} \\
\bottomrule
\end{tabular}}
\caption{Point Cloud and Image Counts by Scene and Period}
\label{tab:pointcloud_image_counts}
\end{table*}

To study the impact of anchor feature initialization, we evaluate three
alternatives to the feature initialization: uniform sampling,
Gaussian sampling, and zero initialization. As shown in
\cref{tab:ablation_init}, zero initialization achieves the best performance across all
metrics, indicating that our model can effectively learn discriminative anchor
features even when starting from a fully collapsed initialization. Uniform
initialization provides competitive performance, while Gaussian initialization
shows slightly degraded accuracy. These results demonstrate that our approach is
robust to the choice of initialization and does not rely on carefully crafted
feature priors.

\section{Additional Results on Other Benchmarks}
\label{sec:other_benchmarks}

Besides ChronoScene, we additionally evaluate ChronoGS on three public benchmarks used in prior long-term or continual neural rendering works: WAT\cite{wu2023cl}, CL-Splats\cite{ackermann2025cl}, and NeuSC\cite{lin2023neural} datasets.
These results complement the main-paper evaluation and provide a broader view of generalization under different assumptions about temporal changes.

On the WAT and CL-Splats datasets (Tabs.~\ref{tab:supp_wat},~\ref{tab:supp_clsplats}), ChronoGS significantly outperforms CLNeRF and CL-Splats, which primarily target incremental or relatively mild scene changes. In contrast, ChronoGS is designed for a \textbf{more challenging setting} with large-scale, long-term variations, explaining its stronger performance on these benchmarks.
On the NeuSC benchmark (only the 5Pointz scene is publicly available, Tab.~\ref{tab:supp_5pointz}), NeuSC focuses on scenarios with \textbf{fixed geometry and frequent appearance changes}, while ChronoGS addresses a more complex setting where both geometry and appearance discretely change. As a result, \textbf{ChronoGS operates in a larger and more challenging solution space}, leading to competitive but slightly lower performance under NeuSC's specific assumptions.

\begin{table}[t]
  \centering
  \small
  \setlength{\tabcolsep}{4pt}
  \begin{tabular}{lcc}
    \toprule
    Method & PSNR$\uparrow$ & SSIM$\uparrow$ \\
    \midrule
    NeRF\cite{mildenhall2021nerf} & 21.36 & 0.709 \\
    MEIL-NeRF\cite{chung2025meil} & 24.05 & 0.730 \\
    CLNeRF\cite{cai2023clnerf} & 25.45 & 0.764 \\
    \midrule
    \textbf{ChronoGS} & \textbf{28.31} & \textbf{0.8934} \\
    \bottomrule
  \end{tabular}
  \caption{Quantitative comparison on the WAT dataset.}
  \label{tab:supp_wat}
\end{table}

\begin{table}[t]
  \centering
  \small
  \setlength{\tabcolsep}{3.5pt}
  \begin{tabular}{lccc}
    \toprule
    Method & PSNR$\uparrow$ & SSIM$\uparrow$ & LPIPS$\downarrow$ \\
    \midrule
    GsEditor\cite{chen2024gaussianeditor} & 24.133 & 0.867 & 0.143 \\
    CLNeRF\cite{cai2023clnerf} & 24.541 & 0.658 & 0.373 \\
    CL-NeRF\cite{wu2023cl} & 23.268 & 0.725 & 0.290 \\
    CL-Splats\cite{ackermann2025cl} & 28.249 & 0.930 & 0.065 \\
    \midrule
    \textbf{ChronoGS} & \textbf{32.03} & \textbf{0.935} & \textbf{0.0431} \\
    \bottomrule
  \end{tabular}
  \caption{Quantitative comparison on the CL-Splats real-world benchmark.}
  \label{tab:supp_clsplats}
\end{table}

\begin{table}[t]
  \centering
  \small
  \setlength{\tabcolsep}{3.5pt}
  \begin{tabular}{lccc}
    \toprule
    Method & PSNR$\uparrow$ & SSIM$\uparrow$ & LPIPS$\downarrow$ \\
    \midrule
    NeRFW\cite{martin2021nerf} & 17.52 & 0.545 & 0.500 \\
    HaNeRF\cite{chen2022hallucinated} & 16.82 & 0.539 & 0.508 \\
    NeRFW-T & 19.41 & 0.611 & 0.418 \\
    HaNeRF-T & 17.90 & 0.585 & 0.445 \\
    NeuSC & \textbf{21.32} & \textbf{0.745} & \textbf{0.274} \\
    \midrule
    ChronoGS & 19.35 & 0.717 & 0.2761 \\
    \bottomrule
  \end{tabular}
  \caption{Quantitative comparison on the NeuSC\cite{lin2023neural} dataset (5Pointz scene). The suffix ``-T'' denotes the variant that uses time as an additional model input.}
  \label{tab:supp_5pointz}
\end{table}


\section{More Qualitative Results}

In this section, we provide additional qualitative comparisons that complement
the results shown in the main paper. First, we include comparisons
with 3DGS\cite{kerbl20233d} and realtime4DGS\cite{yang2023real}, which are omitted from the qualitative figures in
the main paper due to space constraints. These results further demonstrate the
advantages of our method in handling multi-period geometry and appearance
variations.

\paragraph{Image Demonstrations.}
We additionally present figures that contain representative qualitative
results for every scene. These visualizations provide a
complete overview of reconstruction quality and temporal consistency across the entire dataset. Please refer to
Figure~\ref{fig:qual_real} and Figure~\ref{fig:qual_synth} for the qualitative results.

\paragraph{Video Demonstrations.}
To provide a more intuitive comparison between our method and static or dynamic baselines, we also include video demonstrations in our supplementary material. These videos show how our method faithfully reconstructs complex geometry and appearance changes across periods. However, static methods, when trained on multi-period datasets that contain both geometric and appearance changes, tend to produce blurred reconstructions in the varying regions and fail to model the distinct scene states of each period. Dynamic methods, on the other hand, rely on assumptions of temporal continuity; when applied to discrete periods, they often yield poor intermediate states or introduce significant artifacts.

\begin{figure*}[h]
    \centering
    \includegraphics[width=\textwidth]{figs/result_real.pdf}
    \caption{\textbf{Qualitative comparison on real scenes of ChronoScene.}
      We compare ChronoGS with representative baselines. 
Each row shows novel-view renderings of different methods, along with ground truth. 
Static models trained on mixed multi-period data produce ghosting and appearance blending due to inconsistent geometry, 
while dynamic methods that assume continuous motion fail under the large temporal gaps, causing incorrect geometry interpolation. 
\textbf{ChronoGS} faithfully reconstructs period-specific geometry and appearance, preserving sharp, temporally consistent details across long-term changes.}
\vspace{0em}
    \label{fig:qual_real}
\end{figure*}

\begin{figure*}[h]
    \centering
    \includegraphics[width=\textwidth]{figs/result_syth.pdf}
    \caption{\textbf{Qualitative comparison on synthetic scenes of ChronoScene.}
      We compare ChronoGS with representative baselines. 
Each row shows novel-view renderings of different methods, along with ground truth. 
Static models trained on mixed multi-period data produce ghosting and appearance blending due to inconsistent geometry, 
while dynamic methods that assume continuous motion fail under the large temporal gaps, causing incorrect geometry interpolation. 
\textbf{ChronoGS} faithfully reconstructs period-specific geometry and appearance, preserving sharp, temporally consistent details across long-term changes.}
\vspace{0em}
    \label{fig:qual_synth}
\end{figure*}

\section{More Information About ChronoScene}

To provide a comprehensive overview of the ChronoScene dataset, we present detailed statistics and visualizations that characterize the temporal distribution of both images and point clouds across different scenes. \cref{tab:pointcloud_image_counts} summarizes the quantitative distribution of images and point clouds for each scene across different periods, revealing the scale and temporal distributions of our dataset. 

\begin{figure*}[h]
    \centering
    \includegraphics[width=\textwidth]{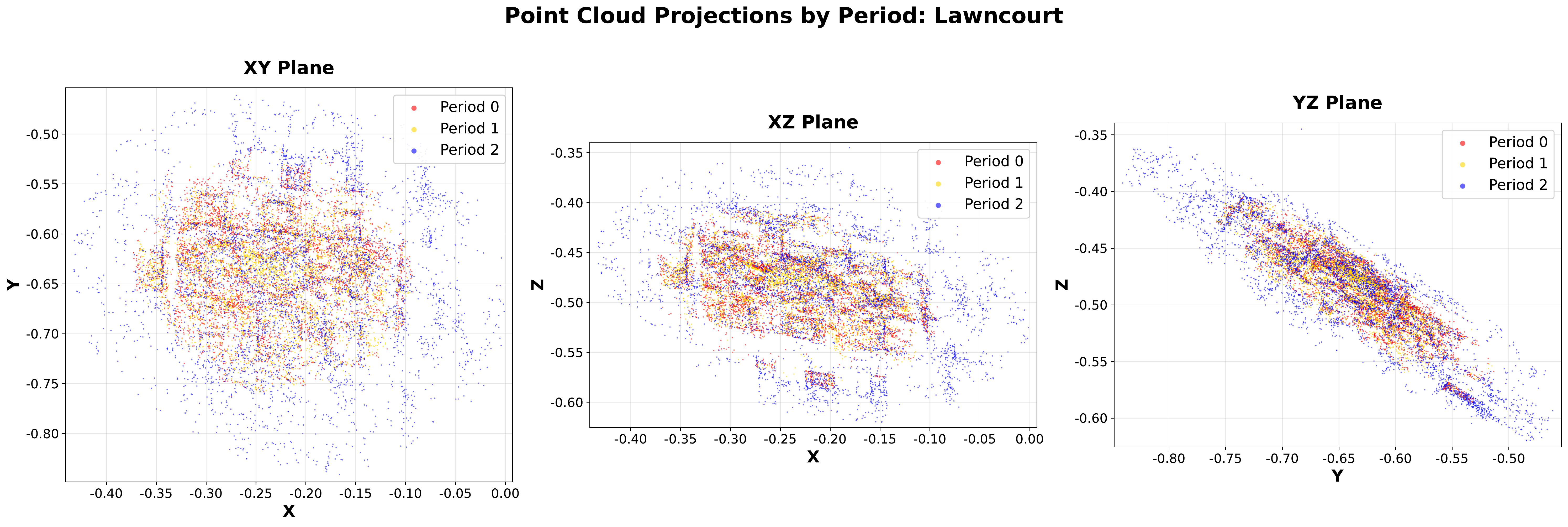}
    \caption{\textbf{Point cloud distribution of scene \textit{Lawncourt}.}}
\vspace{0em}
    \label{fig:pc1}
\end{figure*}

\begin{figure*}[h]
    \centering
    \includegraphics[width=\textwidth]{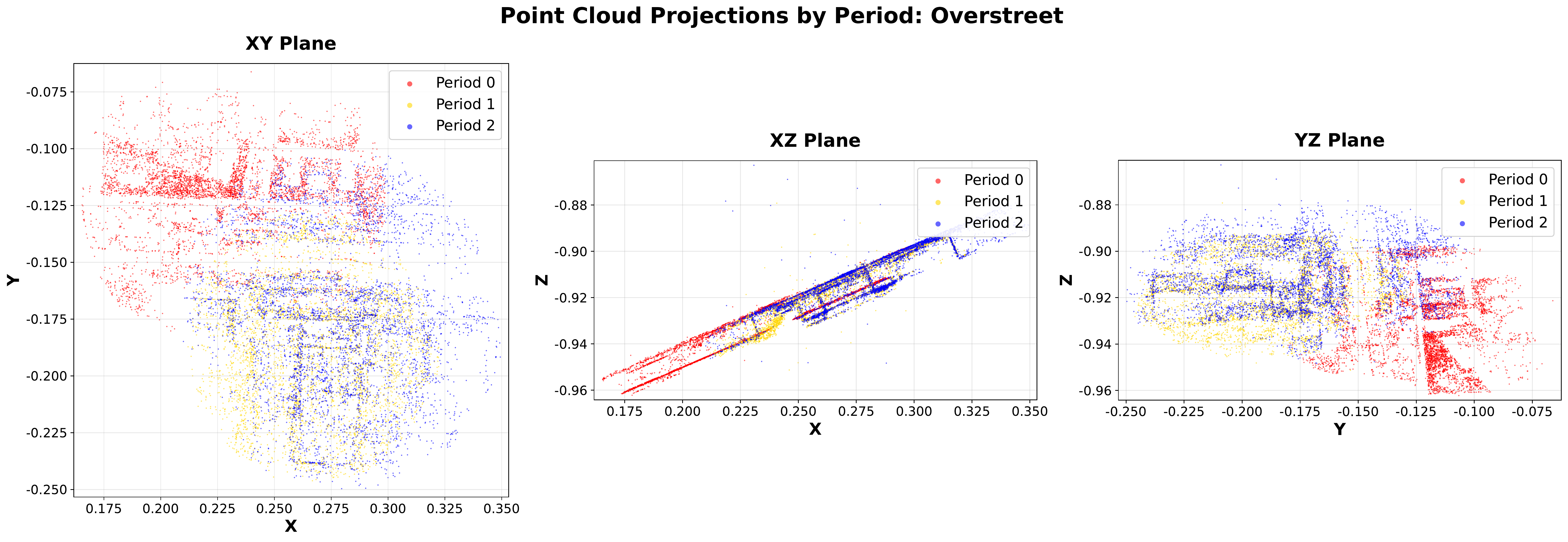}
    \caption{\textbf{Point cloud distribution of scene \textit{Overstreet}.}}
\vspace{0em}
    \label{fig:pc2}
\end{figure*}

\begin{figure*}[h]
    \centering
    \includegraphics[width=\textwidth]{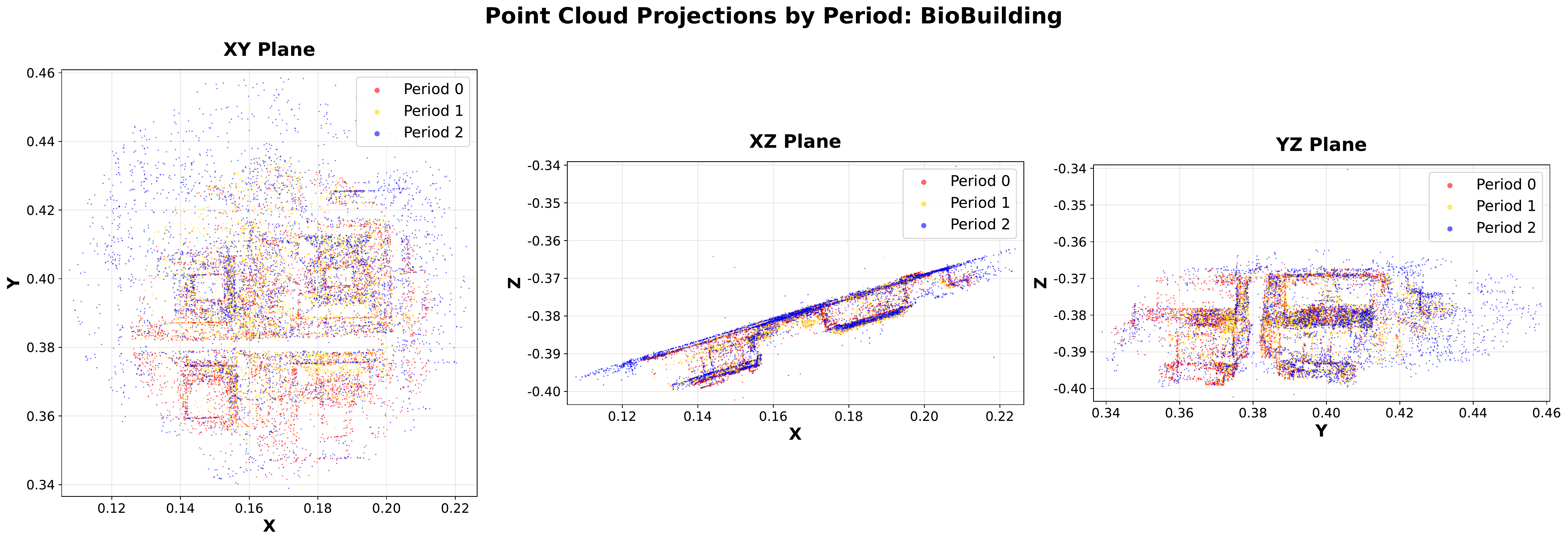}
    \caption{\textbf{Point cloud distribution of scene \textit{BioBuilding}.}}
\vspace{0em}
    \label{fig:pc3}
\end{figure*}

\begin{figure*}[h]
    \centering
    \includegraphics[width=\textwidth]{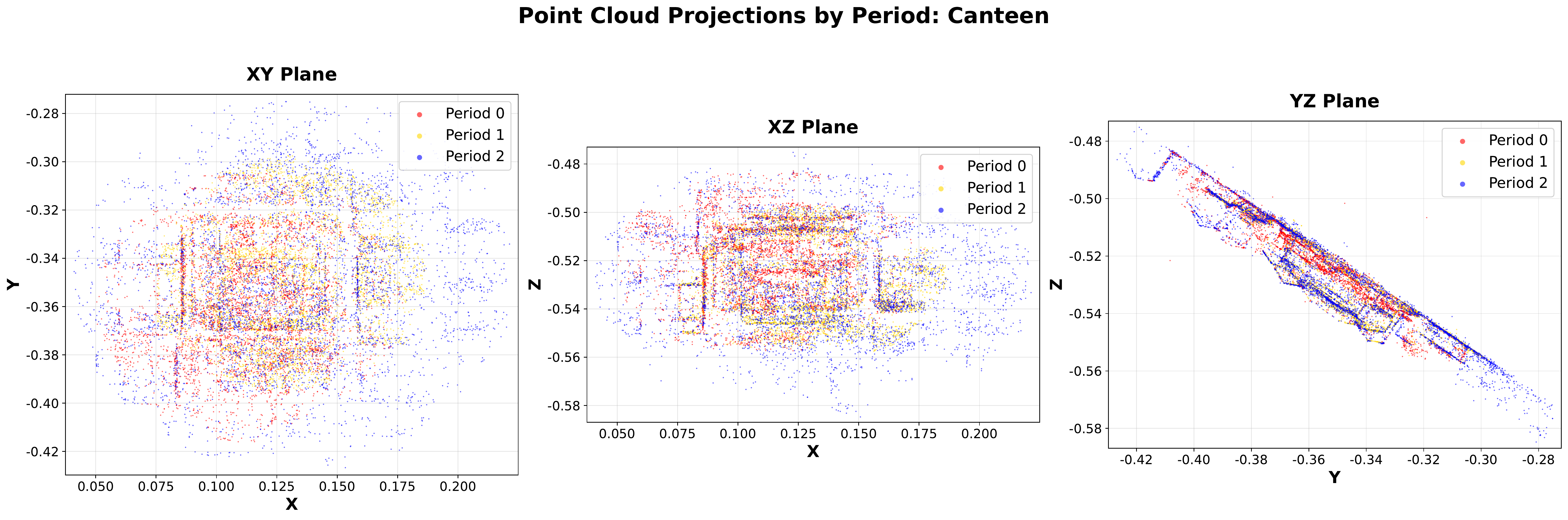}
    \caption{\textbf{Point cloud distribution of scene \textit{Canteen}.}}
\vspace{0em}
    \label{fig:pc4}
\end{figure*}

\begin{figure*}[h]
    \centering
    \includegraphics[width=\textwidth]{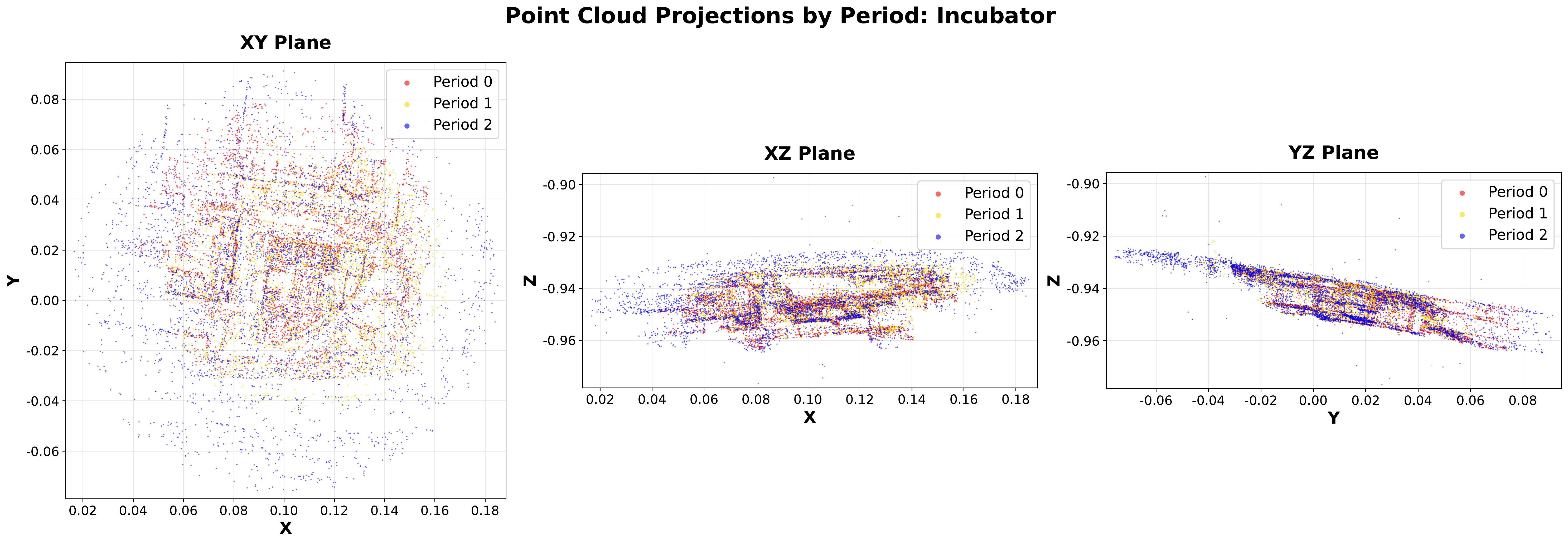}
    \caption{\textbf{Point cloud distribution of scene \textit{Incubator}.}}
\vspace{0em}
    \label{fig:pc5}
\end{figure*}

\begin{figure*}[h]
    \centering
    \includegraphics[width=\textwidth]{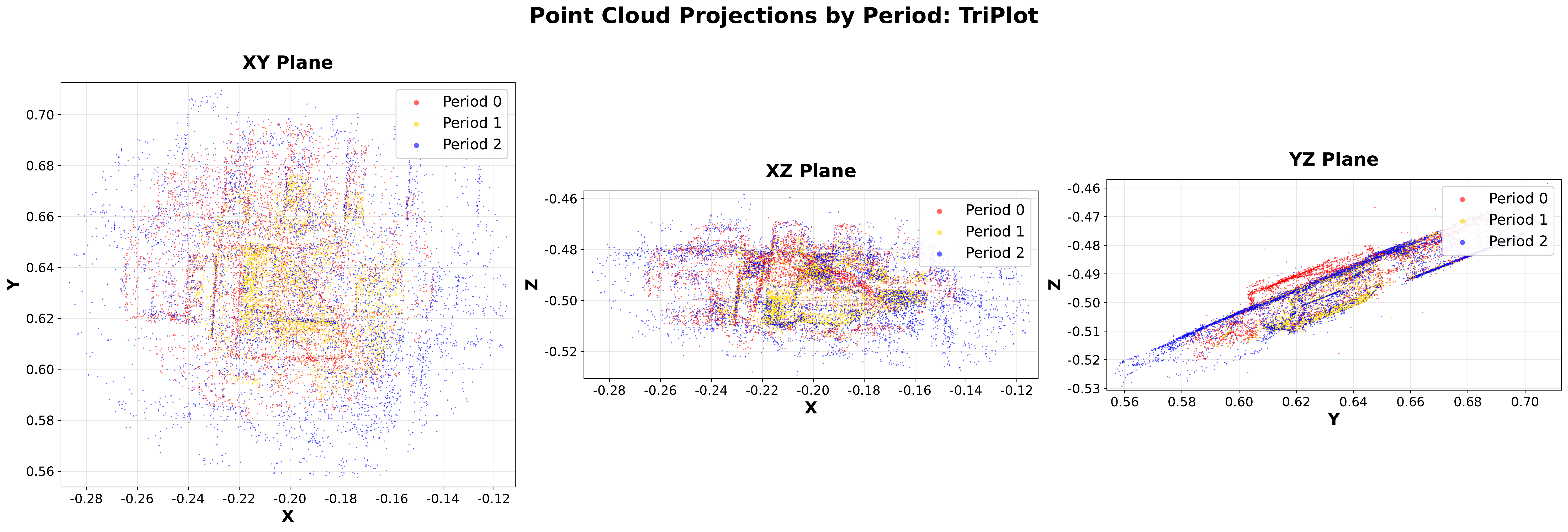}
    \caption{\textbf{Point cloud distribution of scene \textit{TriPlot}.}}
\vspace{0em}
    \label{fig:pc6}
\end{figure*}

\begin{figure*}[h]
    \centering
    \includegraphics[width=\textwidth]{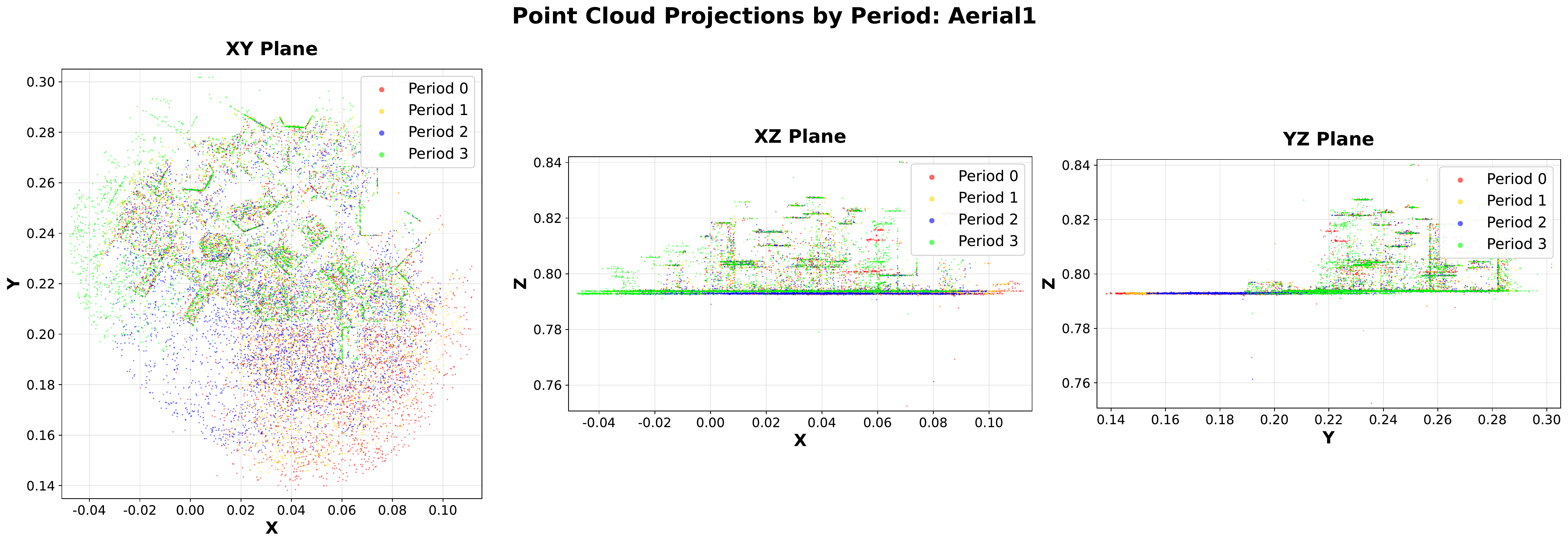}
    \caption{\textbf{Point cloud distribution of scene \textit{Aerial1}.}}
\vspace{0em}
    \label{fig:pc7}
\end{figure*}

\begin{figure*}[h]
    \centering
    \includegraphics[width=\textwidth]{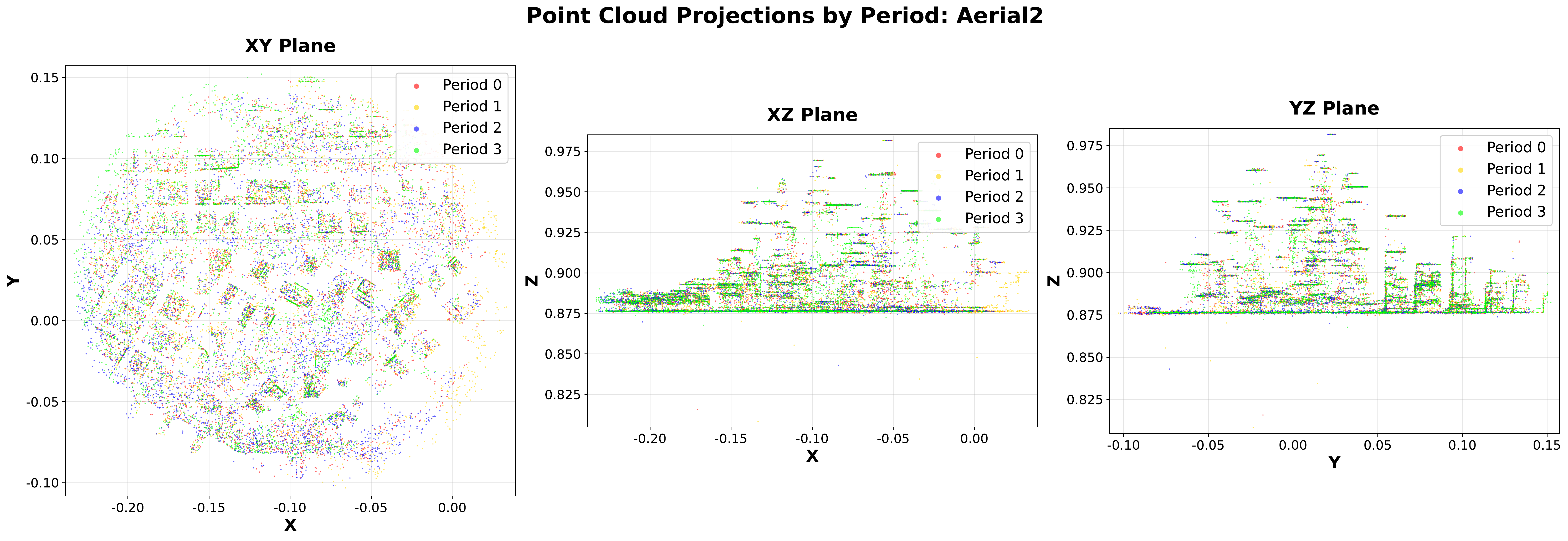}
    \caption{\textbf{Point cloud distribution of scene \textit{Aerial2}.}}
\vspace{0em}
    \label{fig:pc8}
\end{figure*}

\begin{figure*}[h]
    \centering
    \includegraphics[width=\textwidth]{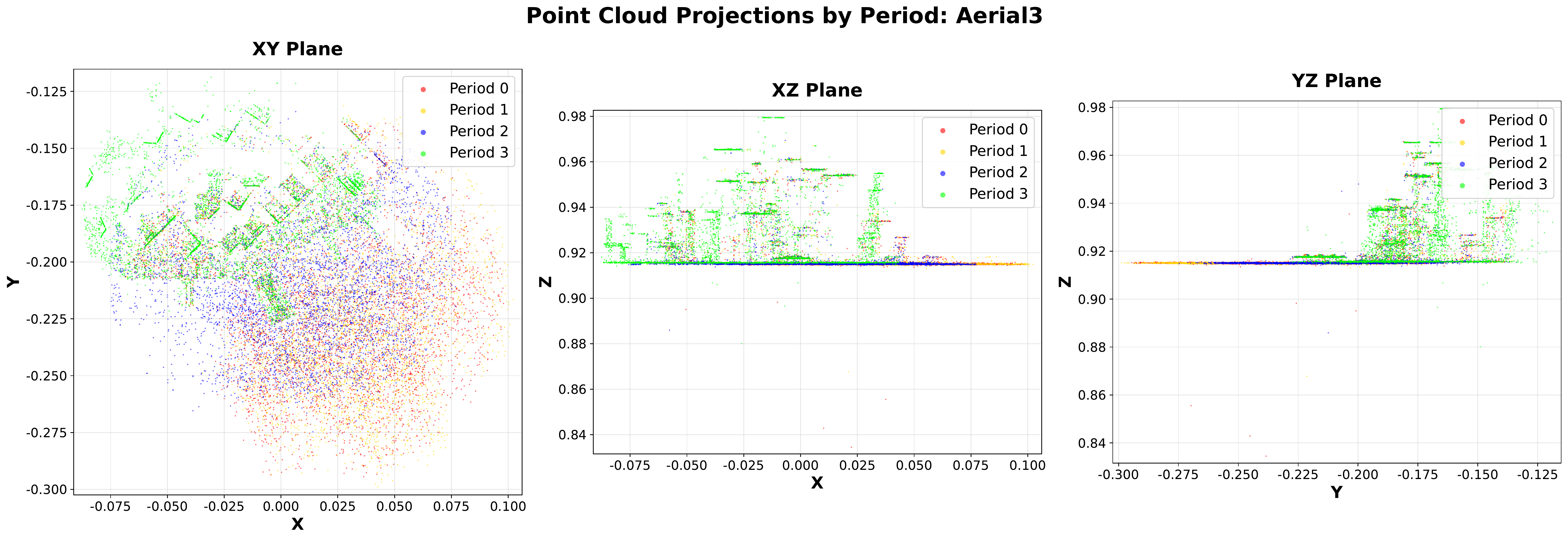}
    \caption{\textbf{Point cloud distribution of scene \textit{Aerial3}.}}
\vspace{0em}
    \label{fig:pc9}
\end{figure*}

\begin{figure*}[h]
    \centering
    \includegraphics[width=\textwidth]{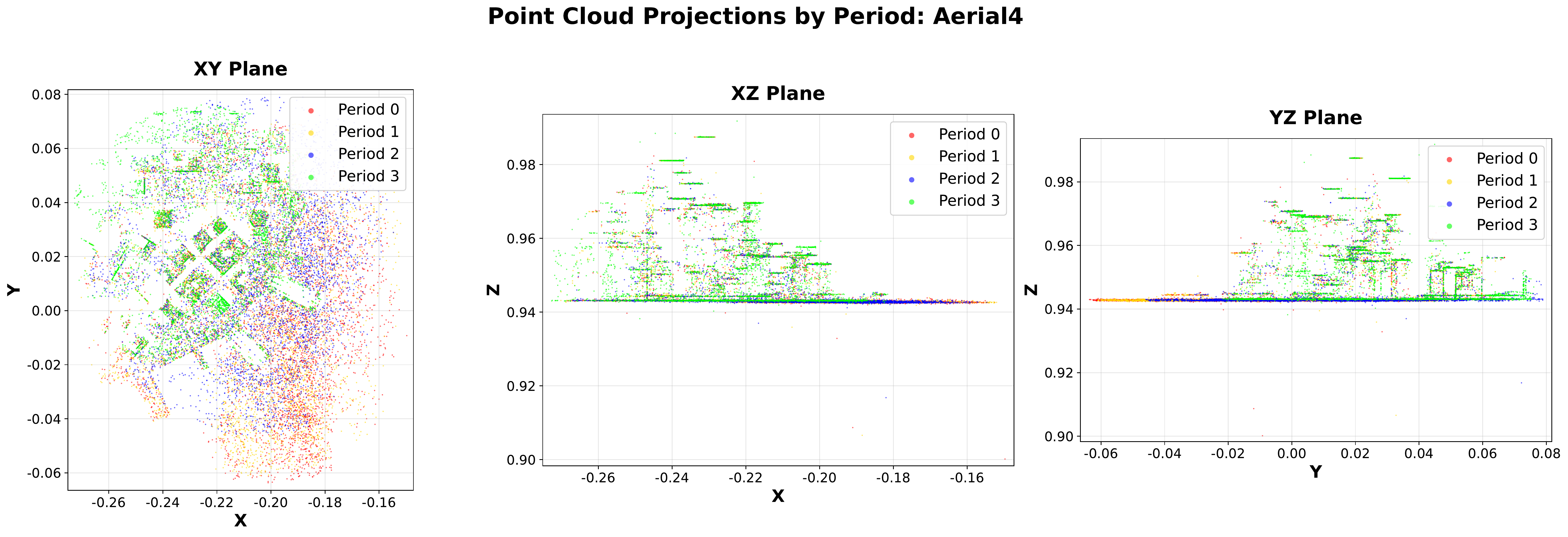}
    \caption{\textbf{Point cloud distribution of scene \textit{Aerial4}.}}
\vspace{0em}
    \label{fig:pc10}
\end{figure*}

\begin{figure*}[h]
    \centering
    \includegraphics[width=\textwidth]{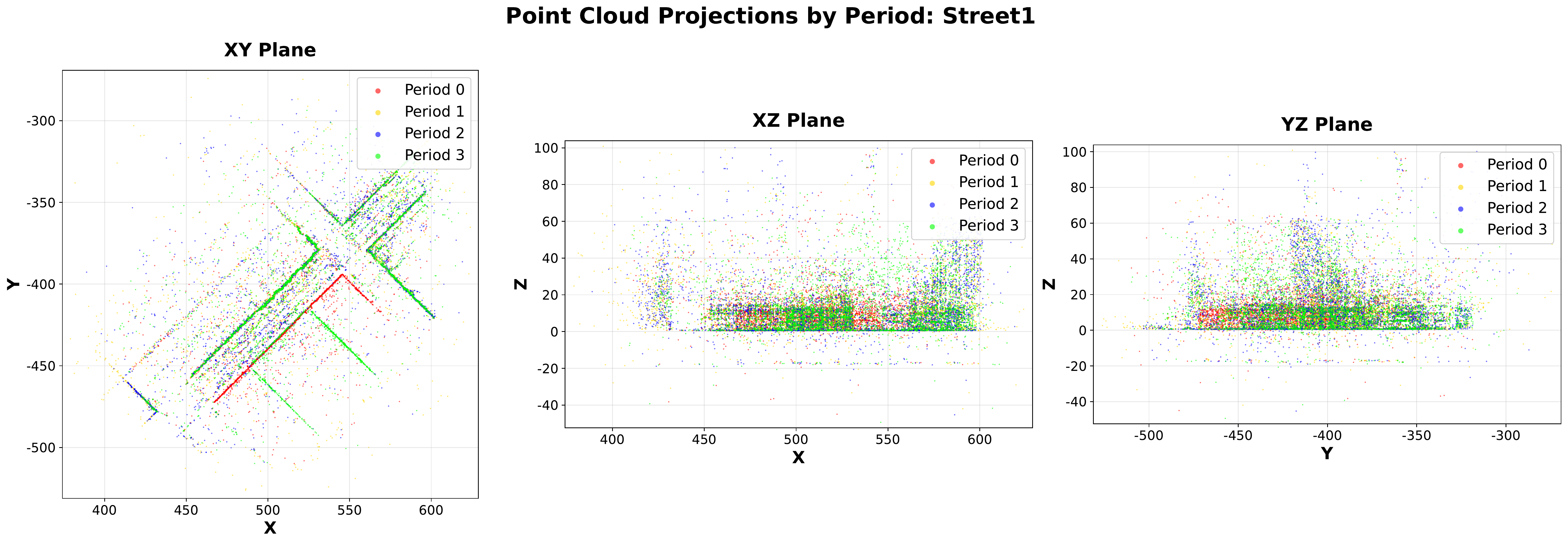}
    \caption{\textbf{Point cloud distribution of scene \textit{Street1}.}}
\vspace{0em}
    \label{fig:pc11}
\end{figure*}

\begin{figure*}[h]
    \centering
    \includegraphics[width=\textwidth]{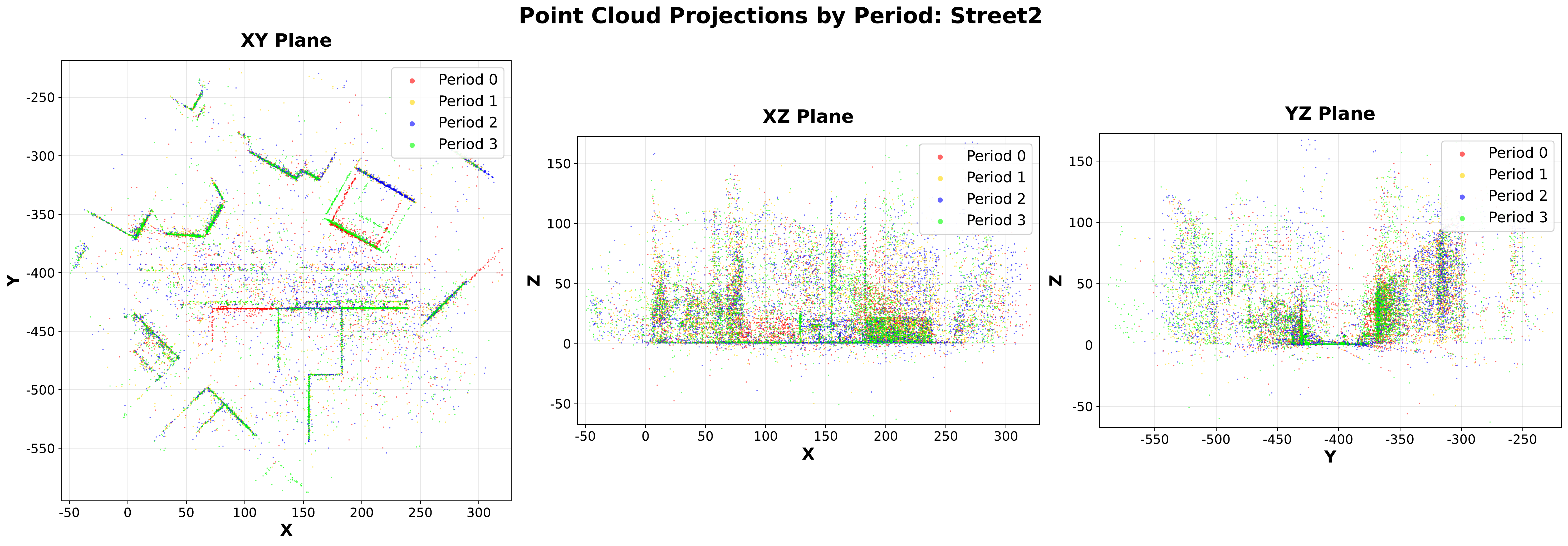}
    \caption{\textbf{Point cloud distribution of scene \textit{Street2}.}}
\vspace{0em}
    \label{fig:pc12}
\end{figure*}

To further illustrate the spatial-temporal structure of the point cloud data, we visualize the \textbf{sparse point cloud distributions} projected onto three orthogonal planes (XY, XZ, and YZ) for each scene, as shown in \cref{fig:pc1,fig:pc2,fig:pc3,fig:pc4,fig:pc5,fig:pc6,fig:pc7,fig:pc8,fig:pc9,fig:pc10,fig:pc11,fig:pc12,}. These visualizations reveal how point clouds are distributed across different periods within the same coordinate system, with each period represented by a distinct color (red, yellow, blue, and green for periods 0, 1, 2, and 3, respectively). The projections highlight both the spatial coverage and temporal variations in the point cloud density. These figures demonstrate that our dataset encompasses a diverse range of scenes with varying temporal characteristics, from street scenes to aerial views.

These analyses collectively demonstrate that ChronoScene provides a rich and diverse dataset for temporal 3D scene understanding, with balanced temporal distributions and comprehensive spatial coverage across multiple scene types.

{
    \small
    \putbib[main]
}

\end{bibunit}

\end{document}